\documentclass[acmtog]{acmart}

\usepackage{booktabs} 

\usepackage{amsmath}
\usepackage{amsthm}
\usepackage{eucal}
\usepackage{graphicx}

\newcommand{\cal}[1]{
  \mathcal{{#1}}
}

\newtheorem{prop}{Proposition}

\usepackage[ruled]{algorithm2e} 

\SetAlFnt{\small}
\SetAlCapFnt{\small}
\SetAlCapNameFnt{\small}
\SetAlCapHSkip{0pt}
\IncMargin{-\parindent}

\setcopyright{none}

\setcitestyle{square}

\begin{document}

\title{Structurally optimized shells}

\author{Francisca Gil-Ureta}
\affiliation{
 \institution{New York University}
}
\email{gilureta@cs.nyu.edu}
\author{Nico Pietroni}
\affiliation{
  \institution{University of Technology Sydney}
}
\email{nico.pietroni@uts.edu.au}
\author{Denis Zorin}
\affiliation{%
  \institution{New York University}
}
\email{dzorin@cs.nyu.edu}

\renewcommand\shortauthors{Gil-Ureta, F. et al}

\begin{abstract}
Shells, i.e.,  objects made of a thin layer of material following a surface, are among the most common structures in use. They are highly efficient, in terms of material required to maintain strength, but also prone to deformation and failure.  We introduce an efficient method for reinforcing shells, that is, adding material to the shell to increase its resilience to external loads. Our goal is to produce a reinforcement structure of minimal weight.  It has been demonstrated that optimal reinforcement structures may be qualitatively different, depending on external loads and surface shape. In some cases, it naturally consists of discrete protruding ribs; in other cases, a smooth shell thickness variation allows to save more material. 

Most previously proposed solutions, starting from classical Michell trusses, are not able to handle a full range of shells (e.g., are restricted to self-supporting structures) or are unable to reproduce this range of behaviors, resulting in suboptimal structures. 

We propose a new method that works for any input surface with any load configurations, taking into account both in-plane (tensile/compression) and out-of-plane (bending) forces. By using a more precise volume model, we are capable of producing optimized structures with the full range of qualitative behaviors.  Our method includes new algorithms for determining the layout of reinforcement structure elements, and an efficient algorithm to optimize their shape, minimizing a non-linear non-convex functional at a fraction of the cost and with better optimality compared to standard solvers. 

We demonstrate the optimization results for a variety of shapes, and the improvements it yields in the strength of 3D-printed objects. 

\end{abstract}

%
%


%
%



\newcommand{\be}{\mathbf{e}}
\newcommand{\bq}{\mathbf{q}}
\newcommand{\ba}{\mathbf{a}}
\newcommand{\bg}{\mathbf{g}}
\newcommand{\bd}{\mathbf{d}}
\newcommand{\bp}{\mathbf{p}}
\newcommand{\bhe}{\hat{\mathbf{e}}}
\newcommand{\bhn}{\hat{\mathbf{n}}}
\newcommand{\bn}{\mathbf{n}}
\newcommand{\bv}{\mathbf{v}}
\newcommand{\bbf}{\mathbf{f}}
\newcommand{\eps}{\varepsilon}
\newcommand{\bu}{\mathbf{u}}
\newcommand{\vone}{\mathbf{1}}
\newcommand{\cE}{\mathcal E}

\graphicspath{{./img}}

\maketitle

\section{Introduction}
\label{sec:intro}

\begin{figure}[!htb]
  \includegraphics[width=\columnwidth]{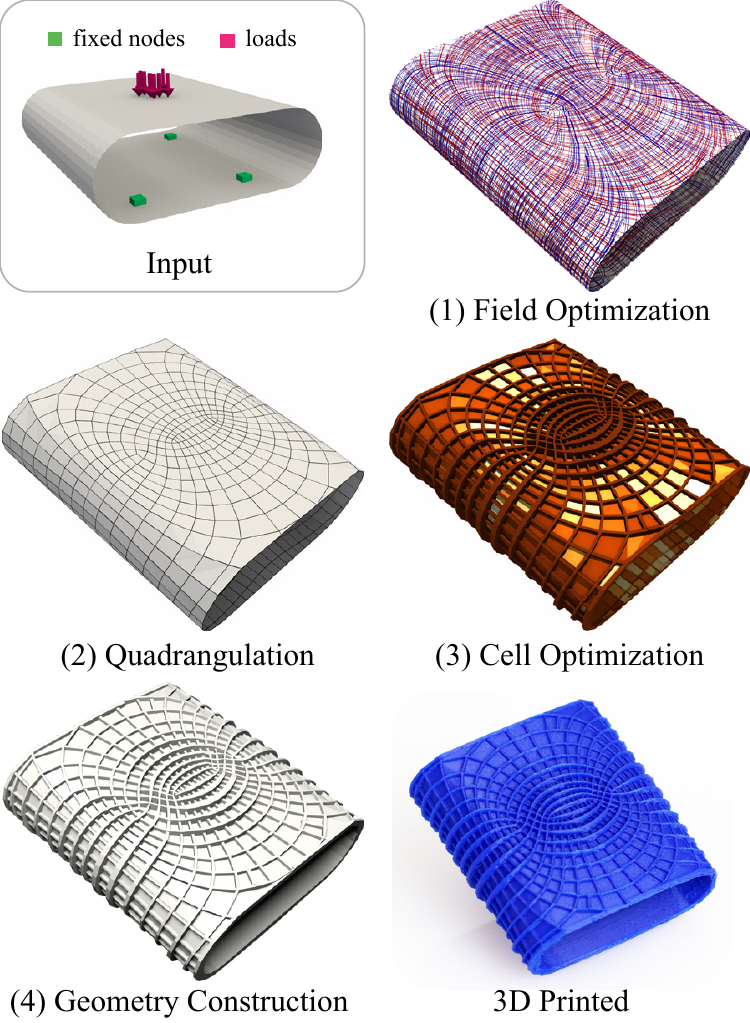}
  \centering
 \caption{Stages of our shell structure generation pipeline.}
  \label{fig:teaser}
\end{figure}
In structural design, shells are considered to be one of the most efficient 
structures because they can be simultaneously lightweight and robust. Shells are 
common in additive fabrication applications because using shells instead of solids 
reduces the cost of material and decreases the fabrication time.  

An optimally-shaped  shell can carry its load relying only on 
tensile/compression forces, with no bending involved, which is very efficient 
in terms of the required material. These types of  shells are commonly found in architecture (domes).  However, the shape of the shell may be determined by considerations other than its  load-carrying properties. For example, the shape of the airplane is determined by 
its aerodynamics; the shape of the car body both by the aerodynamics and aesthetics; the top of
a table or a shelf is flat, as objects need to be placed on it; the artistic intent primarily determines the shape of a lamp or a statue. 

\begin{figure}[!htb]
  \includegraphics[width=\columnwidth]{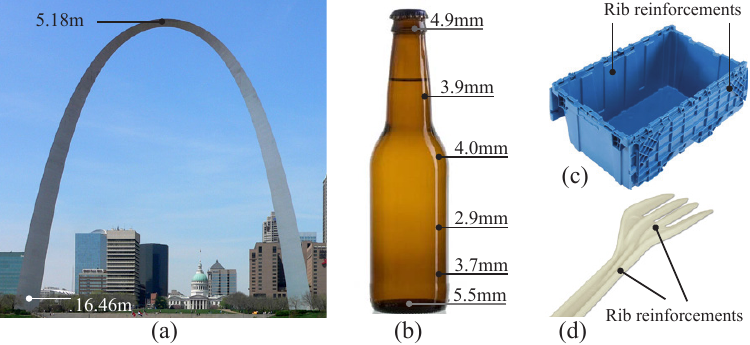}
  \centering
  \caption{Examples of shell structures showing variable thickness walls and 
            ribs: (a) Gateway Arch monument; (b) beer bottle with standard 
            thicknesses; (c) plastic shipping container; (d) plastic fork.
  } 
  \label{fig:shell-examples}
\end{figure}

Shell structures with shapes fixed by considerations other than loading are often 
reinforced by additional means, most commonly increasing thickness in critical 
areas or adding ribs (Fig.~\ref{fig:shell-examples}).  Formally, a common 
optimization goal for a shell reinforcement structure is to \emph{minimize the 
weight of the added material while keeping the maximal stress of the structure 
bounded}. The first ensures the structure remains lightweight, while the second 
prevents structural failure.

This problem has been well studied for two 
dimensions in the limit of low volumes. In 2D, the optimal layout forms a pattern of orthogonal lines (Hencky-Prandtl net) and, for a given 
layout, the minimum weight structure has all members fully stressed. These structures 
form,  in the limit of low volumes, classical Michell structures and can be 
obtained by solving a convex optimization problem.
It is also well understood for pure bending problems for plates, i.e., the special 
case of flat shells with loads orthogonal to the surface; in this case, it also 
reduces to a different convex problem. 

The situation is far more complex for the reinforcement of fixed curved shells 
embedded in three dimensions that we consider.  For these shells, the weight-optimal structure may  be locally either beam-like, forming \emph{ribs} aligned with stress directions,  or membrane-like, forming  variable-thickness walls with no perforation
\cite{sigmund2016non}. The first case typically corresponds to bending-dominated regions while the second to 
areas dominated by in-plane forces.  The optimal local structure is determined by the surface shape,  the supports, and the loads. 

In this general case, the problem is no longer convex and cannot be optimally 
solved either by methods that assume that result is only a variable thickness 
shell or by Michell-truss type methods. 

In this paper, we propose a novel efficient computational method for constructing 
optimized reinforcement structures for shells naturally producing a full range of 
behaviors spanning the space between variable-thickness shell and rib-type 
reinforcement.

We partition the problem into three steps: (1) determine the field of (approximately) 
optimal stress directions; (2) construct the skeleton of the reinforcement structure 
that follows these directions, forming polygonal (predominantly) 
quad cells aligned with the field; (3) optimize how material is distributed inside 
the cells.

\paragraph{Contributions.}
\begin{itemize}
\item For computing the field of optimal stress directions, we developed a generalization of Hencky-Prandtl nets which takes bending into account and can still be solved by minimizing a convex energy.

\item For material distribution optimization, we use a low-parametric structure model for cells to efficiently  optimize the distribution of the material.  As the global  optimization problem is  \emph{fundamentally non-convex} (we discuss the reasons on Section~\ref{sec:motivating}), to solve it we propose an efficient global/local method which shows stable and fast  convergence behavior.  
\end{itemize}

We demonstrate our approach by optimizing several 3D shapes. This evaluation 
shows our method can handle shells with arbitrary curvature, and successfully 
transitions between membrane- and bending-dominated regions, obtaining the expected 
optimal substructures.

\section{Related Work}
\label{sec:related}

Our work builds on the ideas from classical structure design for 2D elasticity 
and plates, with the key ones originating the work of Michell \cite{michell1904l}.

We complement these fundamental ideas with quadrangulation techniques which can 
be reinterpreted as a way to transition from an infinite continuum of 
field-aligned beams to its discretization. We use a variation of 
\cite{bommes2009mixed}, but any conforming, field-aligned method could be used 
(e.g., \cite{kalberer2007quadcover, Myles:2014,Aigerman:2015,CampenBK15,esck2016}). 
The optimization method for computing the optimal strain field can be viewed as 
a specialized cross-field optimization method. Similarly to the recently proposed 
method of \cite{Knoppel:2015} it has the advantage of being convex.

\paragraph*{Shell Optimization}
The closest works to ours are the recent works \cite{kilian2017material} and 
\cite{li2017rib}, with which we share a number of ideas. 
The former describes an
elegant connection between curvature and Michell trusses and optimizes the surface shape so principal stress and curvature directions  coincide. Only tensile forces are considered, and the volume approximation they use is valid for narrow beams (see Section~\ref{sec:motivating}). Similar to our work, \cite{li2017rib} keeps the shell surface fixed. This  work considers a network of ribs, aligned with stress lines, and minimizes their  volume;  similarly to \cite{kilian2017material}, this work uses a narrow-beam approximation for the volume, and always produces a thin-beam structure. The cross-section shape of individual beams is optimized, which produces additional weight reduction. We discuss differences to these works in more detail in Section~\ref{sec:evaluation}.

Our approach also shares similarity with \cite{groen2017homogenization},  which uses similar structures for constructing 2D optimized structures. 

On the other extreme, \cite{zhao2017stress} considers optimization of variable 
shell thickness, while keeping the topology fixed, which is suboptimal for bending 
reinforcement. Our work aims to bridge the gap between these extremes. 

The approach of \cite{pietroni2015statics} aligns a network of beams to an input 
stress field. Another recent related work, \cite{jiang2017design} considers 
structures made out of beams with a small number of distinct cross-sections. Both 
methods are suitable for architectural design; instead, we focus on applications, 
like 3D printing, which allow greater flexibility of structures. 

\paragraph*{Structural Optimization}
The literature on structural optimization is quite extensive, and there is no 
chance that we can do justice to all of it. The main types of approaches found in 
the literature include topology optimization methods (SIMP or ESO-based), analytic 
methods for optimal structures directly based on Michell-type theories, and 
methods based on shape derivatives (using an explicit or implicit evolving 
surface representation). Important books, which include reviews of many other 
works are \cite{rozvany1976optimal}, \cite{allaire2002shape}, \cite{bendsoe2013topology}, 
as well as recent reviews, \cite{munk2015topology} and \cite{sigmund2013topology}.

The most prevalent methods in topology optimization of structures are based on 
SIMP-type methods (see \cite{bendsoe2013topology} for a review), which relax the 
problem to optimizing a density over a domain, which is then converted to a 
structure by thresholding. This approach has many advantages, including simplicity 
of implementation \cite{sigmund200199}, connection to homogenization theory, 
flexibility in integrating functionals, and ease of scalable implementation 
(\cite{Aage2015toptop} and \cite{wu2016system}). Nevertheless, the result will 
typically depend on initialization: for complex topology to emerge, the domain 
needs to be discretized using a fine grid.  The parameters of the result 
(e.g., the sparsity of the structure, or minimal thickness) 
need to be controlled indirectly through algorithm parameters. Finally, the result 
is a voxelized structure, which then needs to be converted in some way to a form 
more suitable for manufacturing. In comparison, our method directly produces 
solutions based on a \emph{globally optimal} field (in low-volume limit) and a 
beam skeleton for the optimized structure, which can be directly adjusted by the 
user in a variety of ways (e.g., converted to a spline-based CAD model if desired). 
We compare in more detail in Section~\ref{sec:evaluation}.

Ground structure methods are among the oldest methods for optimizing topology of 
truss/beam structures. These methods start with a structure consisting of a large 
number of redundant beams and optimize it to determine the cross-sections, which 
automatically eliminates some of the beams. Recent examples of applying these type 
of methods include \cite{sokol201199,zegard2014grand,zegard2015grand3}.
Compared to our approach, ground structure methods have to restrict the directions 
of beams to a small set, which affects both optimality and flexibility of the design. 
The larger the initial set of beams, the closer they may approximate the optimal result. 
In computer graphics, the ground structure method was used early in 
\cite{smith2002creating} for truss structure design. \cite{panetta2015elastic} used a 
version of a ground structure method to obtain initial topologies for computing 
microstructures with prescribed material properties, followed by shape optimization.

To a great extent, our work was inspired by the beautiful structures explored in 
the literature on analytic or semi-analytic structure design, e.g. 
\cite{rozvany2012structural}, which includes many examples of exact problem solutions, 
such as Hencky-Prandtl nets. Our goal is to use this type of ideas in the
general setting of surface domains, taking advantage of the optimality criteria
and insights into the structure of the solutions. A concise exposition of the 
theory underlying Michell-type optimal layouts can be found in \cite{strang1983hencky}.
We note that the application of Michell-type structures in 3D is only appropriate 
for certain problem settings: e.g., with no lower-bound constraints on shell thickness, 
variable thickness shells are likely to emerge as a solution \cite{sigmund2016non}.

Shape-derivative based optimization techniques (e.g., \cite{Allaire2008909}) can obtain
very good results when one needs to improve an existing design, by evolving the 
shape to a local minimum. However, while level-set methods of this type allow for 
topology changes, the result does vary considerably depending on the starting point.
In contrast, our goal is to obtain a starting point that is close to the global optimum, 
as long as the desired structure has a relatively low volume. 


\paragraph{Digital fabrication.}
The works closest to ours in this domain are \cite{li2010beam}, \cite{tam2015slam},
and \cite{tam2015principal}. These methods are based on constructing structures 
from stress lines on surfaces which, while different from the optimal fields we 
compute, are often a close approximation.  The overall pipeline of the method of 
\cite{li2010beam} is similar to ours: they start with a field, and construct 
trusses following the field by tracing lines from supports to loads. The method 
is limited to two dimensions and demonstrated only for relatively simple structures. 

\cite{tam2015slam} uses FDM to add material directly along the principal stress 
lines, on 2.5D surfaces. The main problems they solve are stress line generation 
and selection. They minimize strain energy subject to a maximal total print length 
(i.e max material) and a consistent maximum spacing between lines.

Applying topology optimization (SIMP and ground structure methods) to 3D printing 
applications is discussed in  \cite{zegard2016bridging}.

\section{Motivating examples}
\label{sec:motivating}

To motivate our method, we start with simple examples of qualitatively different behavior of optimized structures. 

\begin{figure}[htb]
  \includegraphics[width=\columnwidth]{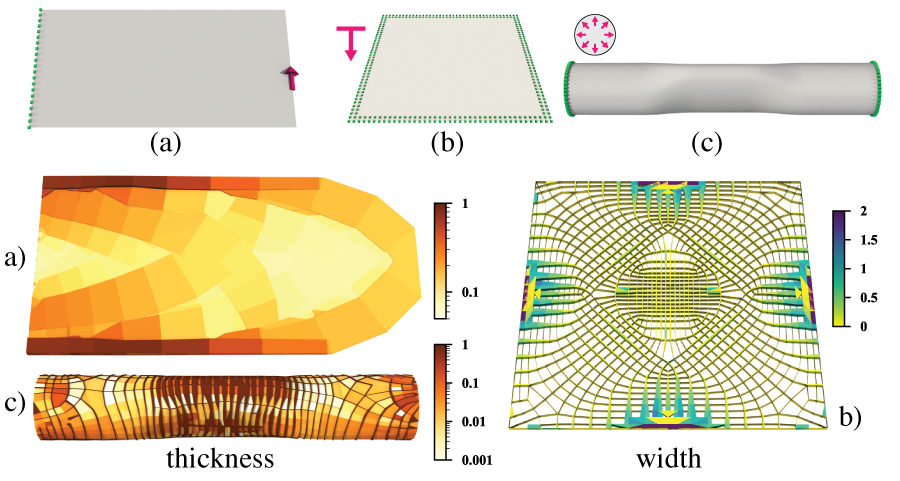}
  \centering
  \caption{(a) optimal structure for a standard cantilever test case with 
    variable thickness, cf. \protect\cite{sigmund2016non}
    (b) bending plate optimization, resulting in a rib structure qualitatively 
    consistent with analytical results; 
    (c) an optimized pipe structure exhibiting a mix of behaviors.
  }
  \label{fig:tensile-bending}
\end{figure}

The two key behaviors of shell-like optimal structures, observed in special-case analytic solutions and topology optimization (cf. \cite{sigmund2016non}), are the formation of discrete narrow \emph{ribs} protruding from the surface in bending-dominated cases (most forces are
perpendicular to the surface), and relatively smooth variation in shell thickness in the pure tensile/compression case (in-plane forces), as shown in Figure~\ref{fig:tensile-bending}.

These behaviors are \emph{not} observed in the simplified models of beam networks approximating a surface: beams in a typical  network do not expand in the direction parallel to the surface, to merge into a variable-thickness shell optimal in such cases.  It turns out that this is due to the qualitatively inaccurate volume computation 
with the volume of the beam network approximated by the sum of individual beam  volumes. We now consider two simple examples showing why this is the case. 

\begin{figure}[t]
  \includegraphics[width=\columnwidth]{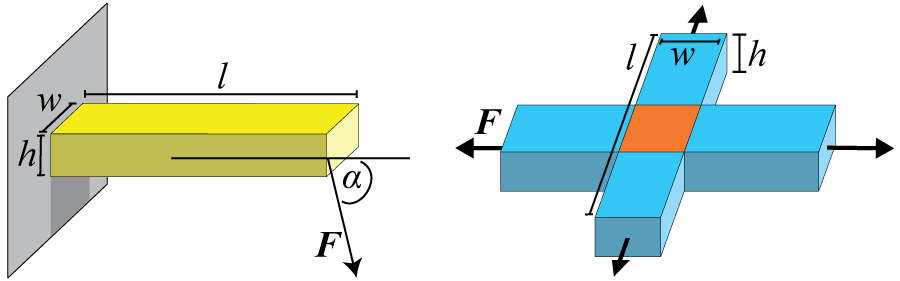}
  \caption{
        Left: beam loading and parameters; Right: surface approximated by two 
        intersecting beams loaded in plane.
  }
  \label{fig:simplebeams}
\end{figure}

First, we consider optimization of a single horizontal beam of width $w$ and height $h$, clamped at one end, and loaded at the other, at an angle $\alpha$ to the beam direction  (Figure~\ref{fig:simplebeams}, left).  This example clarifies optimal behavior  when  there is stress in only one direction, both for bending 
($\alpha = \pi/2$) and tension ($\alpha = 0$). The second example involves two intersecting beams (Figure~\ref{fig:simplebeams}, right). It is a simple model for a piece of a surface  with there is stress in two directions. 

For a single beam, the force along the beam is proportional to  the cross-section $wh$. The force \emph{perpendicular} to the beam  (bending) is proportional  to $wh^3$.  The total force is proportional to  $\cos\alpha wh/l + \sin\alpha wh^3/l^3$, where $l$ is the beam length, 
and $\cos\alpha$ and $\sin\alpha$ are due to projection to the direction of $F$.
For simplicity, assume the proportionality constant to be 1.
We optimize the beam volume $whl$, keeping the force balance constant:  
$\cos\alpha w(h/l) + \sin\alpha w(h^3/l^3) = F$. Eliminating $w$, yields unconstrained  minimization of $V = Fl^2/(\cos\alpha +y^2\sin\alpha)$, with $y=h/l$.  Clearly, the solution is to maximize the relative thickness $y = h/l$, if $\sin\alpha \neq 0 $: increasing thickness has a higher payoff as forces grow as $h^3$ vs. only $h$ for width.
However, in the case of a vertical beam, $\alpha = 0$;  the constraint takes the 
form $wy = F$, i.e., the volume value is fixed and the choice of $y$ makes no  difference. We conclude that the solution can be always taken to be the "thick but narrow"  beams (we refer to them as \emph{ribs}), with maximal  possible $h$. 

The situation is more complicated for surfaces. In the second example, we approximate the surface  locally by beams aligned with the perpendicular stress directions (see figure ~\ref{fig:simplebeams} right).  For simplicity, we assume the 
forces, widths $w$, lengths $l$ and thicknesses $h$ to be the same for both beams.
If we view the intersecting beams individually and approximate the total volume as $V = 2whl= 2wyl^2$, then the reasoning above applies to each beam: 
for $\alpha = 0$, only the cross-section matters, but even a small
bending component will prioritize maximal $h$ solutions, so both beams 
will be thick and narrow; for in-plane forces $F$, we get $wy = F$
and the optimal volume  $V = 2Fl^2$ no matter which $w$ we use. 

Considering beams in separation  ignores the fact that \emph{the intersection area of the beams is counted twice}: 
this part of  material is performing ``double work'', supporting loads in two directions along two beams. 
 The correct combined volume of two beams is given by 
 $$V'(w,y) = 2wyl^2 - w^2yl,$$
 assuming the same beam width and  thickness for both. For in-plane forces, as before, we have the constraint  $wy = F$ for constant $F$.   The functional $V'$ can be expressed then as    $V'(w) = Fl(2l-w)$,
 by eliminating $y$.  From the expression it is clear that one would want to maximize the \emph{width} in this case as opposed to \emph{thickness}. 
The optimal volume for max thickness $w =l$ is $Cl^2$. 
In comparison, if we use large thickness $y = C/w$, then optimal 
$w \approx \epsilon$ is close to zero, and the volume $V'(\epsilon)$ is close to $2Cl^2$, two times higher than optimal $V'(l)$.

In the case of two intersecting beams with an out-of-plane load in addition to in-plane, requiring a combination
of a tensile and bending force to support, there is an optimal trade-off between $w$ and $h$ minimizing the volume, as long as loading cannot be supported by pure bending forces. 

If we further impose constraints on maximal and minimal surface thickness, even in  tensile-dominated areas, beams would form, because the optimal solid shell there would be  too thin. A general optimization method should be able to smooth between grillage-like  structures for bending dominated areas and ``thin and wide'' structures for the rest  of the surface. 

\emph{We conclude, from these examples, that to reinforce a shell in a manner close to optimal for arbitrary loads and shell shape, both solid variable-thickness
and rib-like structures may be required in different areas of the surface, 
and for these to emerge, in a beam-based optimization problem, a non-convex volume function accounting for beam intersection areas has to be used.} 


\section{Problem formulation}

We start with a description of the discretization of variable-thickness perforated surface shell structure that we use, and the optimization problem we aim to solve. 

\subsection{Perforated variable thickness shells}
\label{sec:paramcell}

Our input is a shell $M$ of an initial thickness $h_0$ (constant per triangle) represented by a triangle mesh,  with a vector of external forces $f$ applied to its vertices, and a set of fixed vertices (supports).

We aim to compute an optimized shape which we call a \emph{perforated shell of variable thickness} $M^p$.   $M^p$ consists of  
(a) a partition $\mathcal P$ of the input surface into polygonal faces (typically quads), 
corresponding to 3D \emph{cells}, and 
(b) an extruded shape for each cell, consisting of \emph{blocks} as described below. 

Given an approximate user target for cell size, our goal is to optimize the edge orientation of the cell boundaries, and thicknesses and width of blocks forming each cell to minimize the weight, while maintaining an upper bound on stresses  (calculated using a beam model). $M^p$ can be viewed as the reinforcement structure for $M$.

\begin{figure}[tb]  
    \includegraphics[width=\columnwidth]{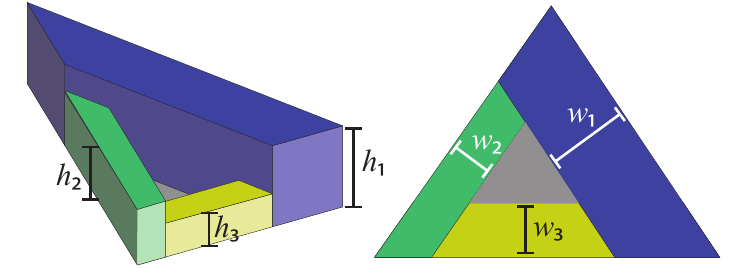}
  \centering
  \caption{Geometric parametrization of triangular cell. Left: perspective view; Right: top view.}
  \label{fig:triangle_volume}
\end{figure}

\paragraph{Cell geometry parametrization.}
For each edge of a face of $M^p$ we introduce two parameters, \emph{width} $w_i$ and \emph{thickness} $h_i$.  With each edge, we associate a hexahedron (block) constructed by creating a strip inside the face at distance $w_i$ from the edge and  extrude the resulting trapezoid along
the triangle normal (See Figure \ref{fig:triangle_volume}). While this geometry may result in gaps between adjacent cells, this has no mechanical implications as we treat each block as anchored to edge vertices.  We model each side of the cell as a beam including tension/compression and bending forces. While this is a very coarse approximation of the shape, it allows us to obtain an approximation of the solution robustly and quickly. This results can be further refined by shape optimization methods (e.g. \cite{panetta2015elastic}).

\paragraph{Volume discretization.}


To simplify our problem, we split all polygonal cells into triangular subcells. 
We refer to the additional edges inserted in this way as \emph{diagonals}. 
We treat these in a special way in the optimization, and in the end ensure that 
the triangular cells can be merged back into the original polygonal cell (Section~\ref{sec:quads}). 

We consider triangular cells with sides $l_i$, $i=1,2,3$, and blocks with rectangular cross-sections of width $w_i$ and thickness $h_i$ along the edge $l_i$  (Figure~\ref{fig:triangle_volume}). 
The simplest approximation of the volume which is typically used in low-volume truss models, when applied to our system would yield simply $\sum_i w_i h_i l_i$.
However, as discussed in Section~\ref{sec:motivating} this approximation 
of the volume results in highly sub-optimal results for shells regions
dominated by tensile forces. 

A more precise approximation  is the volume of the extrusion of 3 trapezoidal regions by different heights.  Denote $y_i = w_i/a_i$, where $a_i$ is the height of the triangle 
with base on side $l_i$, $A = \frac{1}{2}a_i l_i$ for any $i$ is the area of the 
triangle. Let $(i,j,k)$ be the permutation of $(1,2,3)$ for which $h_i \geq h_j \geq h_k$, the volume is given by the  following simple expression: 
\begin{equation}
\begin{split}
V(w,h) = &A \left( (2-y_i)y_i h_i + (2-2y_i-y_j) y_j h_j\right.\\ &\left. + (2-2y_i-2y_j-y_k) y_k h_k\right)
\end{split}
\label{eq:tri-volume}
\end{equation}
This volume can be written as $V(w,h) = \max_{(i,j,k)} V_{ijk}(w,h),$ 
where $V_{ijk}(w,h)$ is given by \eqref{eq:tri-volume} for arbitrary permutation 
$(i,j,k)$.  This expression for the volume is useful for the optimization method 
in Section~\ref{sec:quads}.

The out-of-plane heights can be constrained not to exceed a user-defined value 
$h_{max}$, and the normalized widths $y_i$ are constrained so that the trapezoidal 
areas do not overlap: 
\begin{equation}
    y_1 + y_2 + y_3 \leq 1,\quad y_i \geq 0, \quad 0 \leq h_i \leq h_{max},  \mbox{for $i=1,2,3$}
\label{eq:constraints}
\end{equation}




\subsection{Elastic deformation discretization} 

We model the perforated shell structure as a \emph{beam network:} for each interior edge, there are two beams, corresponding to the blocks of incident cells along the edge. 

\paragraph{Notation.}

The beam network consists of a set of beams $\cE$ that are joined together at 
nodes. For a node $i$, $N(i)$ is the  set of indices of nodes connected to it, and the vector  $\be_{ij}$  connecting nodes $i$ and $j$ corresponds to the edge $e_{ij}$. For each edge,  $\bhe_{ij}$ denotes the unit vector along $\be_{ij}$. The edge $e_{ij}$ has length $\ell_{ij}$. We assume that all cells are made of uniform material with $E$ as Young modulus. We use $\sigma[ij]$ and $\eps[ij]$ notation for the one-dimensional (tensile or bending) stresses and strains  of beam connecting vertices $i$ $j$ in a given cell.

\paragraph{Beam linear elasticity discretization.}
The tensile strain along each beam is the scalar $\eps^t[ij] = (\bu_j-\bu_i)\cdot \bhe_{ij}/l_{ij}$, same for both blocks at 
the edge $e_{ij}$ where  $\bu_i$ is the displacement of a vertex $i$.
It is related to the stress by $\eps^t[ij] = \sigma^t[ij]/E$.

\begin{figure}[!tb]
  \includegraphics[width=0.7\columnwidth]{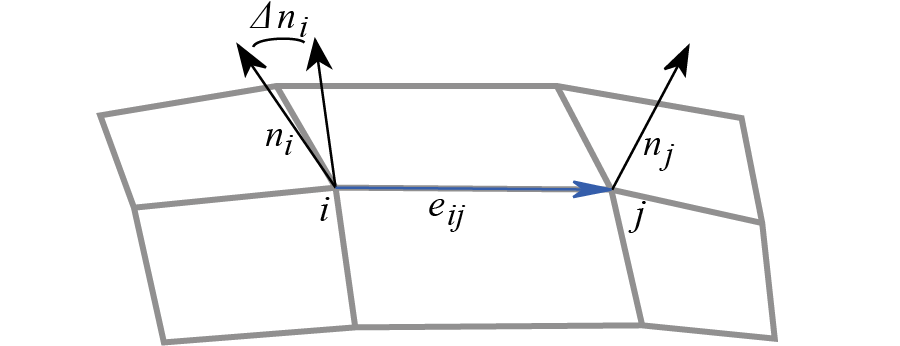}  
  \centering
  \caption{Beam bending discretization.}
  \label{fig:beam-bending}
\end{figure}

For our problem, bending discretization is critical.  We use a pure displacement-based beam bending approximation, but a more standard
beam element could be used.  Our beams are \emph{clamped} to a freely rotating plane at each vertex, i.e., preserve the angle between the beam and the (freely moving) normal to that plane.  We use  a simple discrete beam model for bending: we neglect the torsion and shear and define the bending strain, i.e., the  change in the curvature of a beam $e_{ij}$ as \begin{equation}
\eps^b[ij] =   \bhe_{ij}^T (\Delta \bhn_j - \Delta\bhn_i)/l_{ij},
\end{equation}
where $\bhn_i$ and $\bhn_j$ are normals meeting at nodes $i$ and $j$ (Figure~\ref{fig:beam-bending}),
and $\Delta\bhn$ denotes linearized change of the normal. The normal change, in turn, is expressed in terms of the displacements $u_{i\ell}$, 
$\ell \in N(i)$ of the incident vertices. 

This leads to the expression for a scalar bending strain on beams:
\begin{equation}
  \epsilon[ij]^b = (D^b_{ij})^T u
\end{equation}  
with the expressions for $D^b_{ij}$ given in the Appendix. 

At a distance $z$ from the middle surface of a beam, the strain is given by $\epsilon^t + z\epsilon^b$, where we omit the beam index.
Based on the standard Bernoulli beam assumptions after integration over $z \in [-h/2,h/2]$ the total beam energy can be expressed as  
\[
\frac{1}{2}( w l h u^T (D^t) (D^t)^T u + \frac{1}{6} w h^3l u^T (D^b) (D^b)^T u),
\]
which leads to the following expression for the stiffness matrix of the beam system:

\begin{equation}
K =  w l h (D^t) (D^t)^T + \frac{1}{6}w h^3l (D^b) (D^b)^T 
\end{equation}  

Given the expression for strain $\epsilon^t + z\epsilon^b$, clearly
both strain and stress are maximal on one of the surfaces, 
i.e., for $z=h$ for a given beam.  This leads to the following stress constraint:
\begin{equation}
|(D^t_{ij} + h_{ij} D^b_{ij})^T u| < \sigma_0
\end{equation}


\paragraph{Optimization problem.} We use index $c$ for triangular cells. Let $w$ is the vector of width parameters of all cells, $y$ is the vector of corresponding normalized widths $w_i/a_i$,  $h$ is the vector of all thickness parameters, $H$ the diagonal matrix with thicknesses on the diagonal, and $u$ and $f$ are displacements and forces respectively. Let $\vone$ be the vector of all 
ones. 

We now formulate our general optimization problem: 
\begin{equation}
\begin{split}
 &\min\limits_{h,y,\cal{P}} \sum_{c \in cells} V(h_c, y_c)\quad 
  \mbox{s.t.} K(h,y) u = f,\\
  &|(D^t \pm D^b H)^T u| \leq \sigma_0 \vone,
  \quad 0 \leq h \leq h_{max},\quad y \geq 0,\\
  & y^c_1 +y^c_2 + y^c_3\leq 1,\;   \mbox{for all cells $c$, $i =1,2,3$} \\ 
\end{split}
\label{eq:opt-problem}
\end{equation}
where the absolute value in the stress constraint is taken elementwise, $\vone$ denotes a vector of ones,  and  the minimum is taken over all partitions $\cal P$  of $M$ into triangulated cells. Additionally, we enforce the same thickness on two sides of all cell diagonals. 

As optimization over all possible partitions into cells is an intractable 
problem combining combinatorial and continuous aspects, we use an 
heuristics to fix the partition first, using \emph{beam continuum approximation}
(Section~\ref{sec:frame-continua}). Once $E$ is fixed, we optimize 
with respect to $w$ and $h$ only.

While the above formula volume differs only by a seemingly simple quadratic term from the simplest approximation, this completely changes the behavior of the problems, and, in particular, the behavior of the solvers.  The problem no longer reduces to convex by a change of variables as it is the case for the simplest formula (cf., e.g., \cite{hemp1973optimum}), and a different type
of solvers need to be applied.  In our experiments, commonly used general purpose non-convex solvers converge very slowly and often fail
to make progress. Our solution is described in Section~\ref{sec:quads}.

%



\section{Overview of the approach}
\label{sec:overview}

Our pipeline for solving the optimization problem consists of the following  steps.

\begin{enumerate}
    \item {\bf Field optimization.} Compute a per-triangle cross-field on surface using stress-based 
    optimization (Section~\ref{sec:frame-continua}).
    This field corresponds to an idealized system of densely spaced thin beams (\emph{beam continuum}) with directions chosen to minimize weight. The problem is formulated in terms of displacements and the desired cross-field is obtained from the symmetric strain tensors. This requires solving a convex optimization problem with inequality constraints.
    
    \item {\bf Quadrangulation.} Create a quad-dominant mesh aligned to this cross-field with a user-controlled spacing of edges. This step is done using a version of mixed-integer quadrangulation \cite{bommes2009mixed}, although any quad layout method with field alignment can be used (Section~\ref{sec:quads}).
    The faces of the mesh will correspond to the \emph{cells} of the perforated shell $M^p$. 
    
    \item {\bf Cell optimization.} Optimize shape parameters of the cells (Section~\ref{sec:quads}). This step has a maximal impact on the 
    final outcome.  We introduce a substructure for each cell, with a small number of control parameters defining its shape (widths $w$ and thicknesses $h$ of rectangular beams along each side). We derive the optimal material distribution by efficiently solving a nonlinear, non-convex problem minimizing the total volume of all cells  with respect to $w$ and $h$, while keeping stresses below a user-defined maximum.  To make the problem tractable we defined  an efficient
    local-global optimization method. 

    \item {\bf Final geometry construction.} Finally, we derive the final geometry of $M^p$    according to optimized widths and thickness. We obtain a triangle mesh by performing a sequence of boolean operations between meshes representing the beams. The final watertight and manifold mesh can be directly used for 3D printing or decomposed into cells for FEM analysis.  
\end{enumerate}

In the following sections, we describe the details of the steps of the pipeline.


\section{Weight-minimizing field optimization}
\label{sec:frame-continua}
In this section, we describe our method for constructing a field 
of directions on the surface approximating optimal directions for weight minimization with bounded stress. The cells in our construction 
will be aligned with these directions. 

The key idea is to solve a version of the beam weight minimization 
we assume that there is a \emph{continuum} of infinitely thin 
infinitely close beams in two orthogonal directions forming the 
surface. The density of the beams and their orientations are 
the optimization variables.  The idea of using this type of continua goes back to   \cite{michell1904l}.

In the simplest case of planar elasticity (which is largely equivalent
to the case of a fixed thickness shell in this case) 
the problem was extensively studied and was shown to reduce 
to a convex problem.  We first describe this classical theory 
(Michell continua) and then extend it to the case of shells with bending
forces (Section~\ref{sec:frame-continua}).


\subsection{Michell continua}


 Here, we briefly review the classical solution, following \cite{strang1983hencky}. The best  directions are known to be the principal stress directions of the \emph{optimized}  structure. (These are \emph{not} the same as the stress direction on the original  shell, although these fields are often close.) 


The force balance for a plate or a shell with no bending is given by the standard  equations in terms of in-plane stress tensor  $\sigma$, strain  $\eps$, possibly varying   elasticity tensor $E(\bp)$, and external force density $\bbf$:

\begin{equation}
  \mathrm{div} \sigma = \bbf,\; 
  \sigma = E(\bp):\eps;\; 
  \eps = \frac{1}{2}(\nabla \bu + \nabla^T \bu).
\label{eq:elasticity-plane}
\end{equation}
where $A:B = \sum A_{ijkl}B_{kl}$ for a 4-tensor and a 2-tensor. 


A   Michell continuum  is an idealization of a beam network, characterized, at every  point, by beam densities $\rho_1$ and $\rho_2$ in two directions, in other words, how many beams cross a unit-length segment along one of the coordinate directions.   In the limit  of small  thicknesses, the total fraction of a small area covered by trusses at a point $p$ is $\rho_1(p) + \rho_2(p)$.  The total volume of the trusses in the continuum can be defined as  $\int_\Omega \rho_1 + \rho_2 dA$.  Note that this 
approximation of area covered by trusses suffers from the same 
flaw pointed out in Section~\ref{sec:motivating}.




The optimal trusses have to be oriented along stress directions, and be critically stressed, i.e., all (non-averaged) stresses on the trusses are equal to maximal  stress $\sigma_0$. This leads to the relationship between $\rho_i$ and corresponding averaged principal stress: $\lambda_i(\sigma) = \rho_i \sigma_0$, $i=1,2$, where $\lambda_i(\cdot)$ denotes the 
$i$-th singular value. 


Then, we obtain the following optimization 
problem for the volume, formulated entirely in terms of stresses. 
\begin{equation}
  \mbox{minimize}\;  \int_\Omega |\lambda_1(\sigma)| + |\lambda_2(\sigma)| dA,\; 
  \mbox{subject to}\; \mathrm{div} \sigma = \bbf.
\label{eq:stress-continuum}
\end{equation}
where  $\sigma_i$ are singular values of the stress tensor.
This problem is known to be \emph{convex} \cite{strang1983hencky} (although it is more difficult than the linear programming formulation for a truss network).
Note that principal stress directions are not fixed and are determined by the optimization. We use these directions as the field for orienting 
beams in $M_p$.

The problem \eqref{eq:stress-continuum} has a simple dual (see appendix) of the form 

\begin{equation}
  \mbox{maximize}\;  \int_\Omega \bbf^T\bu dA\; 
  \mbox{subject to}\;  |\lambda_i(\eps)| \leq \eps_0, i=1,2.
\label{eq:dual-continuum}
\end{equation}  
Where $\eps$ is the strain of the optimal solution. The dual problem is significantly easier to deal with in the case of continua.

We note that here we neglect the overlap volume of trusses discussed in 
Section~\ref{sec:motivating}; while it could be included as $-\rho_1 \rho_2$ term,  this would immediately make the problem  non-convex, and the benefit of more precise field optimization in terms of volume reduction 
is minor (Section~\ref{sec:evaluation}). 

\subsection{Continuum optimization with bending}
Next we generalize problem \eqref{eq:dual-continuum}  to include bending forces.   


If the thickness of the shell remains fixed, one can add bending to the functional with relative ease without changing convexity of the problem. We set the shell thickness in this case to half of the maximal allowable thickness; while the resulting field is suboptimal, as  we show experimentally in Section~\ref{sec:evaluation}, inaccuracy in the beam direction has less effect on the overall weight reduction, compared to width/thickness optimization of beams. 

\begin{figure}[t]
  \includegraphics[width=\columnwidth]{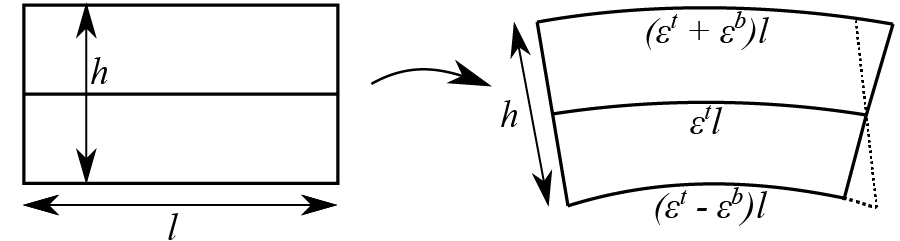}
  \centering
  \caption{Vertical strain distribution in a shell.}
  \label{fig:bending}
\end{figure}

We make the standard assumption of planar stress for the shell, i.e., no stresses are active in the direction perpendicular to the shell surface. The strain at a distance $z$ from the midline of the shell is given by $$\eps(z) = \eps^t + z\eps^b,$$

where $\eps^b$ is the \emph{bending strain} tensor, equal to the linearization of the change in the  shape operator $\nabla \bhn$ (Figure~\ref{fig:bending})

\begin{figure}[b]
  \includegraphics[width=\columnwidth]{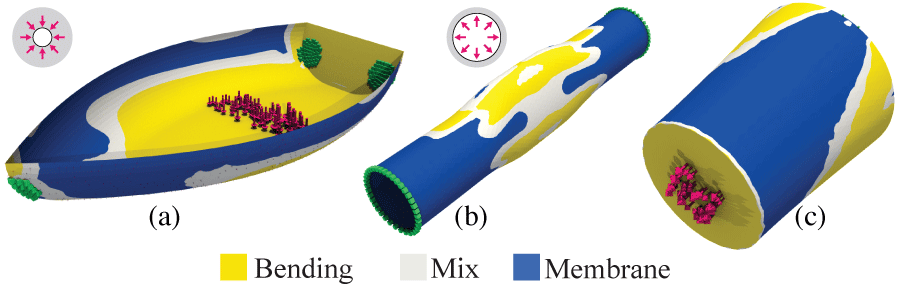}
  \centering
  \caption{Loaded shells dominated by bending and tensile forces, and a mix of these. Support nodes are highlighted in green while loads are shown in red. Loads on (a) and (b) include, respectively, external and internal pressure. }
  \label{fig:shell-types}
\end{figure}

Consistently with the Michell continuum, we seek to minimize the total weight  of a beam continuum, bounding the stress everywhere by $\sigma_0$.
We observe that the eigenvalues of a $2\times 2$ symmetric matrix $A$  using the already mentioned substitutions
$a = (A_{11} +A_{22})/2$, $b = (A_{11}-A_{22})/2$, $c = A_{12}$, are of the form $a \pm \sqrt{b^2+ c^2}$. Note
that these are respectively convex and concave functions of the argument, and, therefore, reach their maxima (respectively minima) on the boundary of the shell, for $z = \pm h$. For this reason, it is sufficient to bound eigenvalues of stress (or strain)  only for $z = h$ and $z=-h$,
to guarantee the bounds elsewhere. 

In the case of bending,  the dual problem formulated in terms of displacements has a simpler form relative to the primal problem: 

\begin{equation}
\mbox{maximize}\;  \int_\Omega \bbf^T\bu dA\; \mbox{subject to}\;  |\lambda_i(\eps^t \pm \frac{h}{2}\eps^b)| \leq \eps_0, i=1,2.
\label{eq:dual-bending}
\end{equation}
Note that now there are two sets of constraints, corresponding to two surfaces of the shell.

The strain tensors $\eps(h/2)$ and $\eps(-h/2)$ can be interpreted as defining cross-fields on the surface.We use the angular average of these fields to align the cells.  
%
Finally, we describe a discretization of this convex problem, and how to use the resulting field to build a mesh. 


\subsection{Discretization and field smoothing}
\label{sec:fields}

 As a first step, we solve a discrete version of the problem \eqref{eq:dual-bending},
which yields displacements $\bu$ at vertices.  From these displacements, we compute the per-triangle strain field eigenvectors, forming a cross-field on the surface, i.e., an assignment of 4 unit vectors, aligned with perpendicular principal strain directions, to each triangle. 

This field is only partially defined: we discuss the properties of the solution that allow us to
identify areas of the mesh where the directions of the field can be used.
In the areas where the field is not defined or is numerically unstable, the beam orientation is not relevant (in the limit of infinite beam density), and we extend the field smoothly. 
We use a cross-field construction procedure of \cite{bommes2009mixed}  to complete the field to the whole surface,
identifying, along the way, the singularities of the field.

\paragraph{Discretization of the optimization problem.}
The optimization problem \eqref{eq:dual-bending} has  a relatively simple discretization that can be readily plugged in into a cone program solver. e.g., \cite{mosek}.

We assume that the surface is given as triangle mesh, $M = (V,E,F)$, and the same notation is used for edge vectors and vertices as we used for beam networks. 

The variables in the problem are displacements, which we discretize using standard piecewise-linear functions on the
surface, with the vector of unknowns $u$ (we use non-bold letters for high-dimensional vectors including all components
of corresponding three-dimensional quantities).

The two quantities that need to be discretized are tensile and bending strains; we define these per triangle.

If $\be_{ij}$ are the vectors along triangle edges, for a triangle $T$, we have the following expression for
the strain, computed as $\frac{1}{2}(\nabla u + \nabla^T u)$:

\begin{equation}
  \eps^t_T = \frac{1}{4A_T}( \sum_{i=1,2,3} \be_{jk}^\perp \bu_i^T  + \bu \be_{jk} ^T)
\label{eq:discr-strain}
\end{equation}  
where $A_T$ is the triangle area,  $j$ is the vertex after  $i$ in CCW order, $k$ is before $i$, and  $\be^\perp_{jk}$
is the vector in the triangle plane perpendicular to the side.

If by abuse of notation, we use
$\epsilon^t_T$  to denote the vector of three distinct components of the strain tensor, we can write
the expression in the form $B_T^t u$, where $u$ is the vector of all displacement degrees of freedom. 

To discretize the bending strain, we use the triangle-based approximation of the shape operator, following
the overall idea in \cite{Onate:1994:STE} and \cite{grinspun2006computing}, using vertices of the edge-adjacent triangles to
compute the changes of the average normals at edge midpoints,  and computing the gradient of the normal using the
formulas \eqref{eq:discr-strain}.

\begin{figure}[!tb]
  \centering
  \includegraphics[width=0.7\columnwidth]{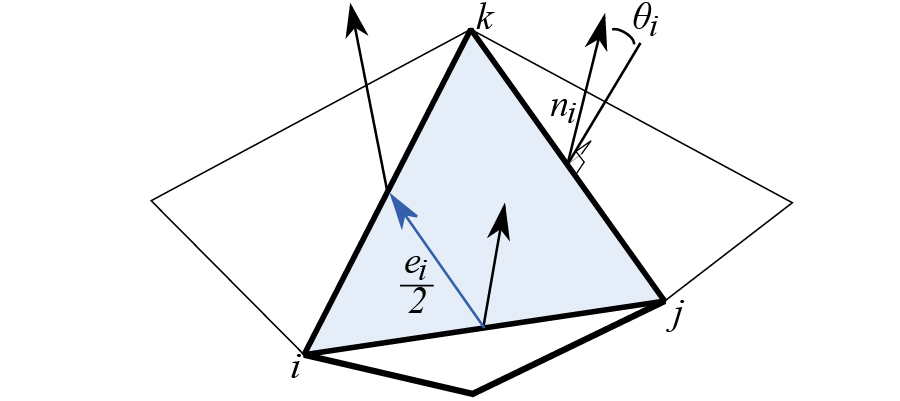}  
  \caption{Discretization of the bending field.}
  \label{fig:discr-bend}
\end{figure}

This leads to the following expressions for the bending strain on triangles, in which we neglect the triangle
deformation: in the deformation modes for which the bending strain is high (i.e., if the curvature changes a lot, in-plane deformations
are small). We use the following formula \cite{grinspun2006computing}, Figure~\ref{fig:discr-bend}:

\begin{equation}
\epsilon^b_T = \sum_{i=1,2,3} \frac{\theta_i}{2A l_i} \be_i^\perp (\be_i^\perp)^T
\end{equation}  
where $\theta_i$ are linearized \emph{changes} in the angles between normals of adjacent triangles. 

Similarly to $\eps^t$, we can write $\eps^b = B^b u$.  Then the discrete problem takes the form

\begin{equation}
\mbox{maximize}\; f^T u\;  \mbox{subject to, for all T,}\;  |\lambda_i(B^t_T \pm h B^b_T)u)| \leq \eps_0, i=1,2.
\label{eq:dual-bending-discr}
\end{equation}

where $f$, similarly to the beam case, denotes the vector of per-vertex forces, and  $u$ is the vector of all vertex
displacements.  We use eigenvalue formulas defined for \eqref{eq:eigenvals} to convert the problem to a
convex cone problem, which we solve using the MOSEK solver \cite{mosek}.

\paragraph{Detecting field zones.} While the output of the previous step defines a tensor for each triangle, not
all of these are meaningful.  In some cases (if the triangle is not deformed at all, or deformed negligibly)
the strain is zero. More generally, some points may have isotropic strains of the form $kI$, where $k$ is a nonzero constant,
for which all vectors are eigenvectors, so the cross-field  is not defined uniquely on this triangle.
For general fields, such points are usually isolated.  However, for the fields corresponding to the solution of the
problem we are considering, the situation is different. There are three possible regimes (see, e.g.,
\cite{strang1983hencky}). 
Specifically, the possibilities include

\begin{enumerate}
\item $|\lambda_1(\eps)| = |\lambda_2(\eps)| = \eps_0$, $\lambda_1(\eps)\lambda_2(\eps) = -\eps_0^2$, principal strains are critical
  and have opposite directions; this corresponds to well-defined two orthogonal beam families; 
\item $|\lambda_1(\eps)| = |\lambda_2(\eps)| = \eps_0$, $\lambda_1(\eps)\lambda_2(\eps) =  \eps_0^2$, principal stresses are critical
  and have the same direction;  in this case, beams are not defined; 
\item $|\lambda_1(\eps)| = \eps_0$, $|\lambda_2(\eps)| < \eps_0$; this corresponds to a single family of beams;
\item $|\lambda_1(\eps)| < \eps_0$, $|\lambda_2(\eps)| < \eps_0$: in this case, stresses (which are dual variables to the
  inequality constraints) are both zero, which means there are no beams in this area.   
\end{enumerate}  

\begin{figure}[!tb]
\includegraphics[width=\columnwidth]{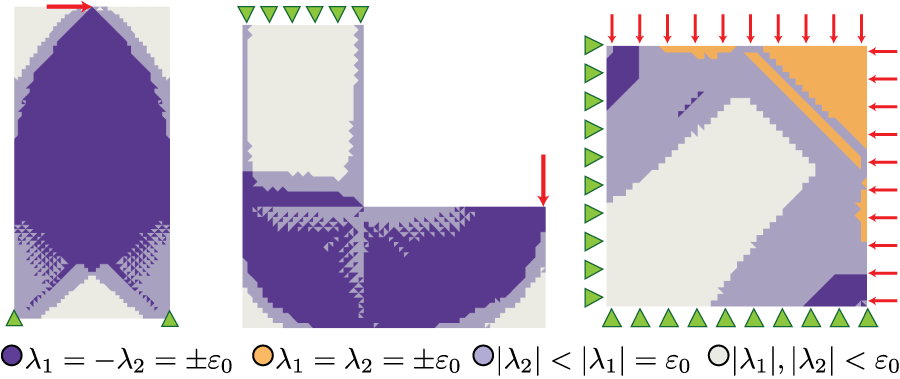}
  \centering
  \caption{Zones of an optimal strain field. The crossfield directions are well-defined only for regions 1, 3 (purple).}
  \label{fig:strain-zones}
\end{figure}

In cases 1 and 3, the crossfield directions are well-defined (purple zones on Fig.~\ref{fig:strain-zones}).  In cases 2 and 4 these are either not defined or are not relevant, due to the absence of structure.  For this reason, for our construction, we use only regions 1, and 3, which we detect by requiring at least one eigenvalue to be close to $\eps_0$, and the difference of eigenvalues to be more than a constant.

We call the resulting field \emph{salient}.  The situation is essentially identical to the cross-fields used for constructing
quadrangulations: typically, a field aligned with curvature directions is used as a starting point, and only in areas
where the difference of two principal curvatures is high. 

\paragraph{Completing the field.} To complete the field on the whole surface, which is needed for a complete structure,
we use cross-field  constrained optimization procedure of \cite{bommes2009mixed}. In this algorithm, the cross-field
is encoded by a per-triangle angle, with respect to a reference direction $\beta_i$ in each triangle. The angles on  salient
triangles are fixed.  On the remaining triangles, these are found by a greedy solve of a mixed-integer problem
minimizing the energy

$$E = \sum_{\mbox{edges}\,(ij)} (\beta_i - \beta_j + k_{ij}\frac{\pi}{2} + \kappa_{ij})^2$$
where the summation is over all dual edges connecting triangles $i$ and $j$, $\kappa_{ij}$ is the angle
between reference directions in triangles, and $k_{ij}$ is an integer unknown accounting for the fact that
cross-field values represented by angles $\beta + k\pi/2 $ are the same.

In the resulting field, defined by the angles $\beta_i$ for all triangles, and integers $k_{ij}$ for
all edges, one can easily compute per-vertex field index and identify field singularities, which become
irregular (valence different from 4) vertices of the quad mesh at the next step. We refer to
\cite{bommes2009mixed} for details of the index computation.

\subsection{Construction of the quad-dominant network}
In general, there may be no optimal beam spacing (in the low-volume limit,
the finer the structure, the lower the optimal volume can be for a given stress). For this reason, the beam spacing is defined
by a user-specified parameter $H$. The most direct approach for constructing a quad mesh aligned with a field would be to trace
it.  However, while it was shown \cite{myles2014robust,ray2014robust} that this approach can be implemented robustly, in general
it requires T-joints (i.e. beams joining other beams in the middle), and it is in general hard to ensure uniform spacing over the
whole mesh. We choose a more conservative approach based on constructing a conforming quadrangulation, without T-joints, 
using  a version of the mixed-integer quadrangulation algorithm \cite{bommes2009mixed} at this step.


While the method does not guarantee perfect alignment of the parametric lines to the input field, it minimizes the deviation in
least-squares sense. We refer to \cite{bommes2009mixed} for further details.

%


\section{Optimizing cell geometry}
\label{sec:quads}

Given an input quad-dominant mesh, split into triangles, we aim at deriving the optimal width and thickness for each edge. As we previously stated, to derive an optimal structure it is important the edges follow the directions we derived in the field optimization step. While our material distribution optimization method works for any mesh, the further the edges deviate from their optimum directions the greater would be the total weight.

\subsection{Optimization algorithm}

We introduce a domain-decomposition-style algorithm for solving the problem  \eqref{eq:opt-problem}.  We observe that in the optimization problem \eqref{eq:opt-problem}, all constraints except $Ku=f$ are localized, 
i.e., each constraint uses variables related to one cell.  Moreover, 
the functional itself is a sum of local volume terms $V^c(w^c,h^c)$.
$Ku =f$ expresses the equation of force balance, i.e., that the sum 
of beam forces at each node is equal to the external force at this 
node.  Our approach is to fix individual beam forces to their values
for current values of geometry parameters $w$ and $h$, and then 
solve for an update to $w$ and $h$ as a set of local volume-minimization problems with variables $w^c, h^c$, replacing the global constraint
$Ku=f$ with local constraints requiring that individual beam forces 
remain the same. 

We start with an outline and then elaborate on how the local step optimization problems are solved.

Initially, we assign sufficiently high values to $w$ and $h$, to ensure that max stress constraints are satisfied. 

\begin{itemize}
\item  {\bf Global step.} The global step is just the standard solve of the elastic equilibrium problem, for fixed cell parameters: \emph{Solve $Ku = f$, for fixed $K$ defined by $w$ and $h$. Compute beam forces as described below.}

\item {\bf Local step.} The local step is the key part of our algorithm. 
Recall that an important feature of optimal structures is that they are \emph{critically stressed} i.e., the maximal stress on any element is equal to the  maximum possible $\sigma_0$.  The reasons for this are straightforward: 
if a stress on an element is below zero, one is free to remove some material, increasing the stress on the remaining part, and decreasing the weight. 

This motivates our approach. For the local step,  we keep the  displacements computed at the global step fixed and solve for widths and thicknesses, that would result in maximal stresses on blocks reaching  the critical value $\sigma_0$ for given displacements.  Each system has 6 unknowns, with 
3 constraints on stresses.

\end{itemize}

\paragraph{Block forces.} To formulate our local algorithm,  we introduce \emph{block forces}, for individual blocks of each cell; we determine 
$w^c$ and $h^c$ for each cell by minimizing the cell volume, while 
keeping the block \emph{critically stressed} i.e., with maximal 
stress $\sigma_0$ and block forces constant. 

Let $K^{loc}$ be the element stiffness matrix corresponding to a block $B$.  The vector of forces corresponding to a beam is the vector  $\nabla_u E^{loc}$, i.e. the derivative of the block energy $E^{loc} = \frac{1}{2}u^T K^{loc} u$ with respect to displacements. Most forces in this vector are zero.  

 We use lower-case, non-bold $d^t$ for the column vector of $(D^t)^T$, and $d^b$ for column vector of $(D^b)^T$, corresponding to the stresses on block $B$.
 After some rearrangement, the force  vector $f^{loc} = K^{loc} u$ due to elastic forces produced by a block is

\begin{equation*}
f^{loc} =  K^{loc} u  =
E w h l \left( (d^t)^T u d^t    + \frac{1}{6} h^2 (d^b)^T u d^b \right)
\end{equation*}
Note that this equation implies that $f$ is in the span of vectors $d^t$ and $d^b$.

Let $\tilde{d}^t$, $\tilde{d}^b$ be the dual pair of vectors to $d^t$, $d^b$.
Let the magnitudes of tensile and bending stresses be $ |E (d^t)^T u| = \sigma^t$, $ |E (d^b)^T u| = \sigma^b$; by taking dot products of both sides with the dual vectors  $\tilde{d}^t$, $\tilde{d}^b$, we arrive at the equation, where we drop beam/cell subsripts:
\begin{equation*}
w h  \sigma^t    = |f^t/l| =  g^t,\; 
\frac{1}{6}w h^3 \sigma^b   =  |f^b/l| =  g^b 
\label{eq:constr}
\end{equation*}

\paragraph{Local optimization problem.}
The stress in the block,  under our assumptions, reaches its maximal value 
at the top or bottom, where its magnitude is equal to $\sigma^t + h\sigma^b$.
This leads to the critical stress constraint  $\sigma^t + h \sigma^b = \sigma_0$.  Using expressions for $g^t$ and $g^b$ above, which we keep fixed at the local step, this  is equivalent to 
\begin{equation*}
 \frac{g^t}{h}  + \frac{6 g^b}{h^3} = \sigma_0 a y
\label{eq:constr-stress}
\end{equation*}
where we have switched to the variable $y=w/a$ introduced in \eqref{eq:opt-problem}, where $a$ is the corresponding cell triangle height.
Without loss of generality, we assume $\sigma_0$ to be 1, which can be achieved by renormalizing all forces to be one. The complete local problem in  variables $h^c_i$, $w^c_i$, $i=1,2,3$ is:

\begin{equation}
\begin{split}
 &\min\limits_{h^c,y^c} V(h^c, y^c)\quad 
  \mbox{s.t.} \quad \frac{g^t_i}{h^c_i}  + \frac{6 g^b_i}{(h^c_i)^3} = \sigma_0 y^c_i a_i\\
  &0 \leq h^c_i \leq h_{max},\quad y^c_i \geq 0,\quad \mbox{for  $i =1,2,3$} \\
  & y^c_1 +y^c_2 + y^c_3\leq 1,\;    
\end{split}
\label{eq:local-opt-problem}
\end{equation}

By eliminating the stresses,\emph{we arrive at a single constraint per block} relating $w$ and $h$, which we express as follows:
\begin{equation}
y =(6 g^b z^2  + g^t z)/a
\end{equation}
where $z = h^{-1}$, the new variable we introduce to simplify the expressions.  This allows us to eliminate
all variables $w^c_i$ from the local optimization problem, leaving only three variables $z^c_i$, with constraints $ 0 \leq z_{min} \leq z^c_i$, $i=1,2,3$, and $z_{min} = (h_{max})^{-1}$.

We say that a cell is \emph{filled}  if the equality $y^c_1 + y^c_2 + y^c_3 = 1$
is satisfied, i.e. the blocks completely fill the cell. 

Without loss of generality, we assume that for the solution 
$z_1 \leq z_2 \leq z_3$; in practice, 6 problems corresponding to 
6 permutations of $(1,2,3)$ need to be solved and minimal solution picked. 

\begin{prop}
The function $V(y^c)$ is a concave function of $y_i$. As a consequence, 
its minima are reached on the boundary of the constraint 
domain; specifically, it is reached at one of the five types 
of configurations: 
\begin{enumerate}
    \item  all three blocks have maximal thickness: $z^c_i = z_{min}$, $i=1,2,3$; 
    \item  the cell is filled, i.e. $y^c_1+y^c_2+y^c_3 = 1$, and no inequality constraint 
    reaches equality; 
    \item the cell is filled, and two thicker beams have equal thickness, 
    $z^c_1 = z^c_2$;
    \item the cell is filled, and  two thinner beams have equal thickness,
    $z^c_2 = z^c_3$;
    \item the cell is filled, and the thickest beam has maximal thickness $z^c_1=z_{min}$
    \end{enumerate}
\label{prop:solution}
\end{prop}  
In the first case, the solution is completely determined. 
In the second case, there are four possibilities: 
no inequality constraint is active (a 2-variable unconstrained optimization problem 
needs to be solved, e.g., parametrized by $z^c_1,z^c_2$); the other  three cases define one-parametric families of solutions, and one-dimensional unconstrained optimization needs to be performed to find exact values, as we explain below. These families can be parametrized by, e.g., $z^c_3=(h^c_3)^{-1}$, with the  values of the remaining $z^c_i$ and $y^c_i$ determined from the active  constraints. The proposition is proved in the appendix.

This behavior of $V$ is in stark contrast to the low-volume formulation ignoring common areas of beam-like parts of the structure: one can see that in three cases out of four, it creates a completely filled cell. 

\paragraph{Solving the optimization problem.} 
Proposition~\ref{prop:solution} leads to an  efficient algorithm for the local step. 

Observe that the constraint $y^c_1+y^c_2+y^c_3=1$ has 
the form 
\begin{equation}
\sum_i g^t_i z^c_i + 6 g^b_i (z^c_i)^2 = 1,
\label{eq:bary-constr}
\end{equation} 
i.e., it is  quadratic in $z_i$. This allows us to reduce the problem to 
a set of unconstrained optimization problems in one or two variables. 

\begin{enumerate}
\item Compute $g^b_i$, $g^t_i$, $i=1,2,3$, for current displacements. 
\item Evaluate $V(z)$, for the case 1 solution with $z_i^c = z_{min}$.
\item For each permutation of $(1,2,3)$ solve three one-dimensional 
optimization problems, minimizing $V(z^c)$, for each of the cases 3-5, and the two-dimensional problem for case 2 
of proposition Prop~\ref{prop:solution}. In each case 3-5, 
substitute the active constraint for $z^c_i$ into \eqref{eq:bary-constr}, 
yielding a quadratic equation in two remaining free variables, 
one of which is $z^c_3$.  Solve it to express the other variable 
in terms of $z^c_3$, and solve a one dimensional optimization problem 
for $V(z^c_3)$. This yields a set of \emph{solution candidates}; 
the minimal solution is guaranteed to belong to it. 
\item Pick the minimal solution from the set of solutions obtained 
for all possible permutations and cases on the previous step. 
\item Update  $w^c_i$ using formula   $w^c_i =(6g^b_i z^c_i  + g^t_i z^c_i)$, and recompute the global stiffness  matrix $K$.
\end{enumerate}  

The convergence behavior of the method is considered in Section~\ref{sec:evaluation}. 

\paragraph{Handling polygonal cells and postprocessing.}  There are several factors not considered
in the solution method above: (1) possible inconsistency of  thickness values across diagonal edges inside triangulated  polygonal cells; 
(2) the coherence of block widths and  thicknesses along the edge lines of the quad mesh, approximating the optimized stress lines. While 
jumps in thickness/width along these lines do not affect the
stresses in our simplified model, in practice, these are likely to lead 
to localized stress concentrations close to jumps, and they are aesthetically 
objectionable. 
(3)  stress values may slightly exceed the  maximal stress after a final global step. 

We experimentally observe that many of the candidate  solutions have close values, especially in areas with no predominant stress direction.

We address (1) primarily in the process of optimization, at step 4, we pick a minimal candidate solution with lowest block thickness on the diagonal, which may not be the most optimal one, as long as it does not deviate above a threshold. Once the optimization is complete, for each subcell of a polygonal cell we increase the lowest block thickness to the maximal minimal thickness over the whole cell.  In addition, for each cell, we store a number of candidate solutions with the smallest volumes.  

We address (2) in a post-processing step using stored candidate solutions: for each non-diagonal edge of a cell, we find its continuation edges along quad mesh edge line in both directions, and  choose the candidate solution closest in width to the average of the previous and next edge widths along the edge line. 

To address (3), we find all blocks with stress exceeding $\sigma_0=1$,  and increase their thickness and width, while maintaining constraints, to decrease the stress to the bound. This process is repeated iteratively until convergence. 
We note that all additional steps are designed to ensure that the final result satisfies stress constraints: in all cases, we never decrease the amount of material in cells, so while the resulting solution may be suboptimal, it always satisfies stress constraints. In practice, the effect of these alterations on the resulting weight is small.

\paragraph{Inflation.} 

\begin{figure}[h!]
  \includegraphics[width=\columnwidth]{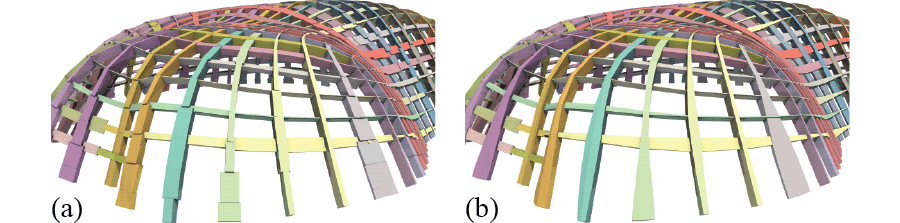}
  \caption{
        Inflation: (a) Each block is meshed independently, leading to visible artifact; (b) An entire stream of edges is meshed to produce a smoother result}  
  \label{fig:meshing}
\end{figure}

Once we have performed the weight optimization, we have one value of thickness and one value of width for each half edge of the optimized polygonal mesh. If we simply extrude the solid block that matches thickness and width for each half edge cell we end up in the situation illustrated in Figure \ref{fig:meshing}.a, where there are visible discontinuities between adjacent blocks, causing possible structural discontinuities. Instead, for each continuous stream of quad edges, we generate a unique solid block that interpolates thickness and width along its length (Fig. \ref{fig:meshing}.b). In detail, given a sequence of aligned edges, we first derive a thickness value for each vertex by averaging the thickness from its adjacent half edges. Similarly, we interpolate widths, but this time we derive two different values for each vertex, one for each side of the sequence. We then define a tangent vector for each vertex as the cross product between its normal and its direction along the edge sequence (obtained by averaging the direction of the two incident edges). Then, having defined a proper reference frame, a thickness, and a width for each vertex, we have all the information to extrude a proper volumetric block. We perform a boolean operation to merge all the blocks together to a manifold watertight mesh using the approach of \cite{Zhou:2016}.


\section{Evaluation}
\label{sec:evaluation}

\begin{figure}[!htb]
\includegraphics[width=\columnwidth]{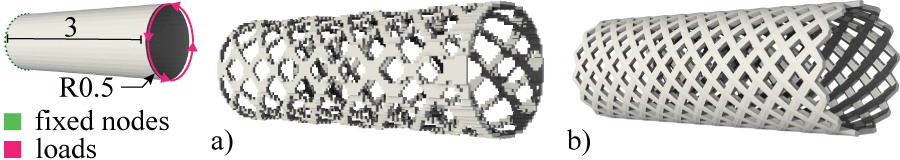}
\centering
  \caption{a) cylinder beam structure obtained using topology optimization with a target volume of $0.375$; b) structure obtained using our method with a maximum thickness of 0.05 and a minimum thickness of $0.04$.}
  \label{fig:torsion-cylinder}
\end{figure}

\paragraph{Topology optimization.} To validate our approach vs. a general-purpose topology optimization method, 
we solve a similar problem with topology optimization code \cite{Aage2015toptop}  by restricting the  
 volume of the material to a cylinder of fixed small thickness, and choosing the volume grid resolution to be half of the cylinder thickness,
leaving little room for shell shape variation.  We observe that for small target volume fractions, as expected, structures emerging in
topology optimization are similar to the beam structures we construct (Figure~\ref{fig:torsion-cylinder}), with similar volume fractions.
For a large volume fraction, the topology optimization method results in variable-thickness shells, but in this case, due to severe
constraints on the thickness of the shell, this does not make a significant difference. Default parameters were used in \cite{Aage2015toptop}.

\begin{figure}[!htb]
\includegraphics[width=\columnwidth]{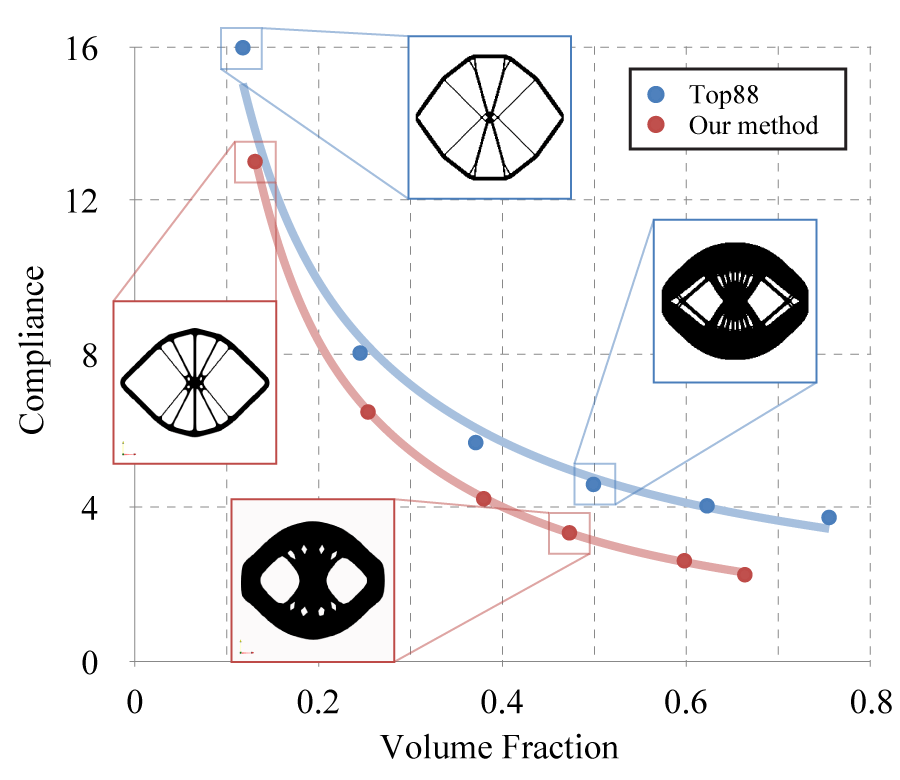}
  \centering
  \caption{Volume fraction vs. compliance energy for a standard example, a cantilever beam,  SIMP topology optimization vs. our method.}
  \label{fig:volume-fraction}
\end{figure}

Figure~\ref{fig:volume-fraction} shows the results of SIMP topology optimization vs. our method, with compliance as a function of the
volume fraction. We observe similar behaviors for both methods. We note that in our case we need to choose the beam spacing parameter:
when this is chosen too coarse, the performance will deteriorate. 

\begin{figure}[b]  
    \includegraphics[width=\columnwidth]{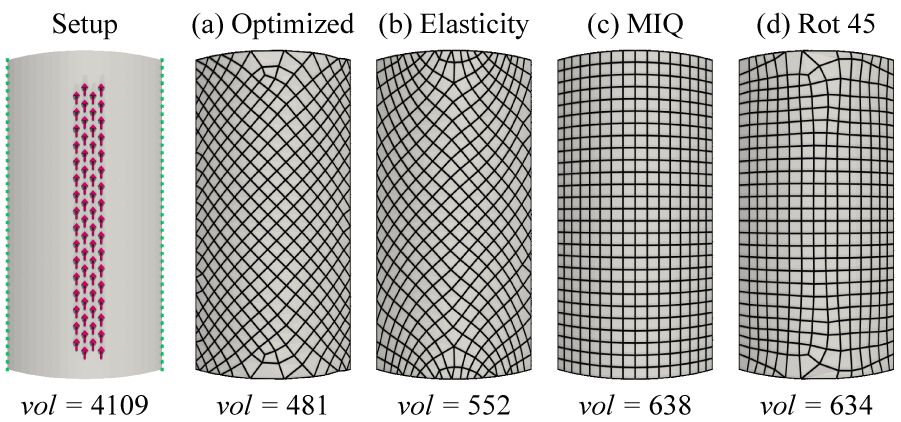}
    \includegraphics[width=\columnwidth]{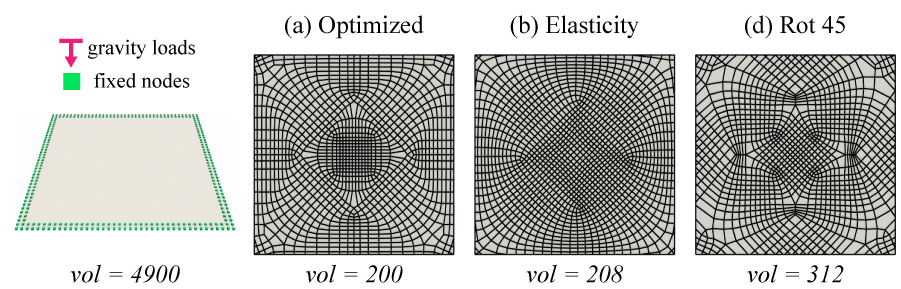}
  \centering
  \caption{Comparison of volumes obtained using different fields for the shell
    structure. Images (a-d) show the quad-mesh obtained with the different fields:
    (a) optimized stress directions from solving Eq.~\ref{eq:dual-bending-discr}, 
    (b) stress directions from solving elasticity, 
    (c) MIQ field constructed using the smoothing method of \protect\cite{bommes2009mixed}, and 
    (d) constructed rotating (a) by 45 degrees.}
  \label{fig:direction-opt}
\end{figure}

\paragraph{The role of beam direction.} In Figure~\ref{fig:direction-opt}, we explore the dependence of the role of beam direction in structure optimality, by comparing structure volume for several fields, in addition to our optimized field (Eq.~\ref{eq:dual-bending-discr}). As a ``worst-case'' baseline
we use the cross-field at a maximal distance from the original field (d);
as one can see, the field makes a significant difference when it is very far
away from the optimal directions, i.e., the choice of directions matters.
Similarly, the curvature field (c), unrelated to stress directions, produces relatively
high values of the volume.  On the other hand, the difference between the optimized
field and the stress field (a and b), while present, is relatively minor. This justifies
the idea to use, as in, e.g., \cite{li2017rib}, the elasticity stress field
instead of the optimized field. The former has the additional advantage of better performance and higher smoothness.

\paragraph{Convergence and dependence on the starting point.}
We have observed that our algorithm almost invariably converges in several iterations, and yields the
expected behavior of solutions in the extreme cases (ribs in the case of bending-dominant shells, and
wide-and-thin beams in high-tension/low-bending areas.   

The plots show the volumes at each iteration for a basic and more complex problem. 
Note that sometimes the optimal value is approached from below: the local step overshoots the volume
reduction and the stress exceeds the maximal allowed level. Nevertheless, the method recovers reliably. 

\begin{figure}[!htb]  
    \includegraphics[width=\columnwidth]{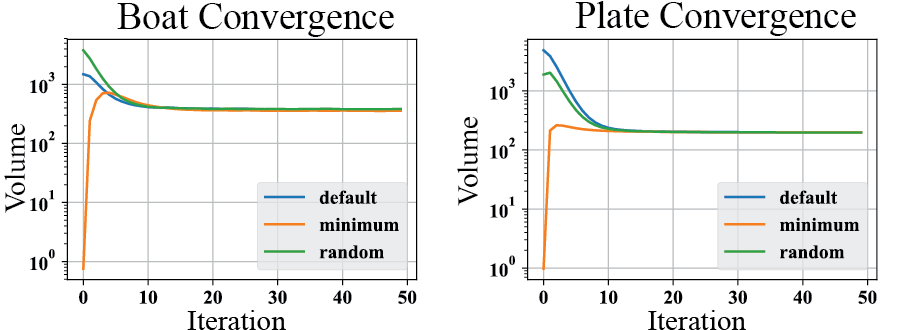}
  \centering
  \caption{Left: Volume convergence for our method, for the boat model, using different starting points.
    Right: similar plots for the bending plate.}
  \label{fig:convergence-chart}
\end{figure}

We also compare our method to two general-purpose constrained optimization methods, SLSQP \cite{kraft1988software} and Ipopt, a barrier interior point method \cite{byrd2000trust}. In this setup, we used an approximation of the beams volume that is smoother (it uses average instead of \emph{max}) and simple lower and upper bounds on the width and thickness. As a starting point, we have used the solution of the convex problem with volume ignoring the overlaps.
Somewhat surprisingly, these methods were not able to change this initial solution by much after a few hundred iterations; although moving in the right direction, in terms of values, it may differ by a factor up to two from the optimal solution.

\begin{figure}[htb]
\includegraphics[width=\columnwidth]{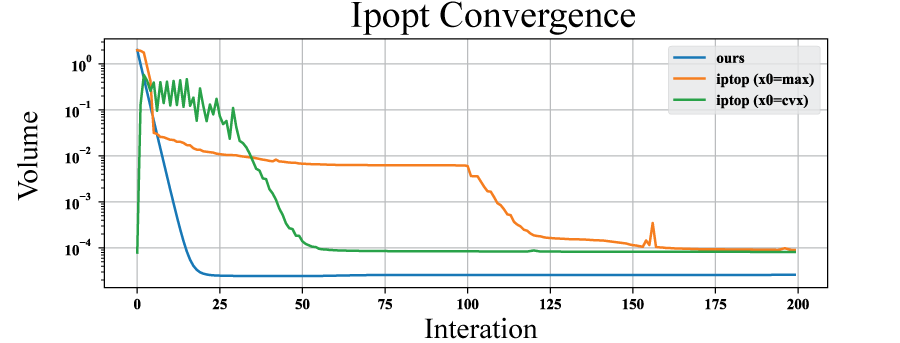}
\centering
  \caption{Volume vs. iteration step for a small cantilever example (408 beams) using Ipopt vs. our method.}
  \label{fig:compare-ipopt}
\end{figure}

While the SLSQP solutions exhibited oscillatory behavior, alternating decreasing the functional with decreasing constraints violations, the interior point method solutions mostly stayed closed to the initial. Figure~\ref{fig:compare-ipopt} shows comparative results with Ipopt for a small cantilever test case, we observe that Ipopt converges to a volume almost an order of magnitude larger.  When Ipopt is initialized with the solution of the convex problem with simplified volume functional (green), the optimization fails to find a better solution. Using a different initialization (orange: maximum width and thickness) does not provide better results. In contrast, our method converges to a much better solution in only a few iterations.

\begin{figure}[!htb]  
  \includegraphics[width=\columnwidth]{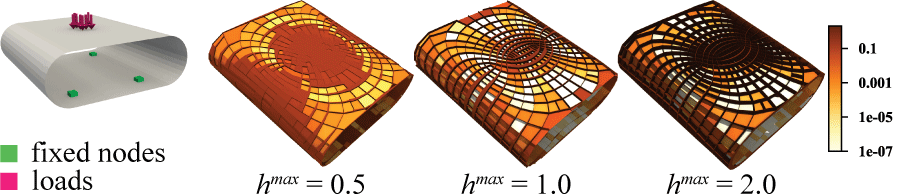}
  \centering
  \caption{In a bending dominated structure, changing the maximal allowed thickness from small to large causes the results to change from a solid plate to a rib-like structure.}
  \label{fig:solid-to-truss-bending}
\end{figure}

\paragraph{Effects of constraints on structure parameters.} Figure~\ref{fig:solid-to-truss-bending} shows 
the effect of increasing maximal allowable thickness in a bending scenario, with the structure moving from 
fully solid, to a structure of narrow but tall beams (if the bound is increased to infinity, in principle, 
any bending force can be realized by a zero-volume infinitely thin rib).

\begin{figure}[!htb]  
  \includegraphics[width=\columnwidth]{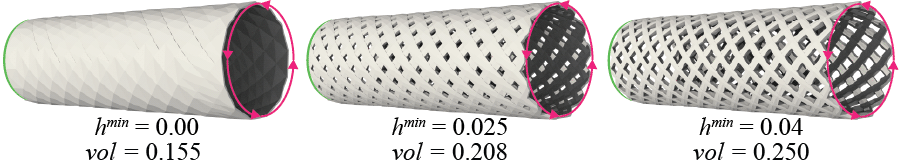}
  \centering
  \caption{In a tensile dominated structure, changing the thickness lower bound from small to large causes the results to change from a solid plate to a rib-like structure.}
  \label{fig:solid-to-truss-tensile}
\end{figure}

Similarly, Figure~\ref{fig:solid-to-truss-tensile} shows the effect of increasing \emph{minimal} thickness 
in a \emph{tensile} scenario. In this case, tall beams appear when increasing the lower bound because 
widths decrease to compensate for the increase of thickness. The usage of beams in tensile scenarios, while 
common, is often a consequence of constraints on minimum element size and are sub-optimal in the absence 
of these constraints. On the torsion cylinder experiment, the effect of using a minimal thickness of 0.04 vs. 0.00 increases the volume by 60\%.

\paragraph{Relation to related work.} A direct apples-to-apples comparison of different methods for optimization for shells is extremely difficult for a number of reasons, most importantly, because of different models used (all approaches use a variation of a beam model, and the specific choice has a significant impact on stress estimation). Another important reason includes a strong dependence of the results on the maximal width/thickness/aspect ratio constraints.  We summarize qualitative differences from the most closely related works \cite{kilian2017material} and  \cite{li2017rib}. 

The key difference from \cite{kilian2017material} is that it changes the shape of the surface, focuses on the case when bending forces can be neglected, and uses the narrow-beam volume approximation. 
This may be appropriate for targeted architectural applications but different types of solutions, produced by our method, may be more optimal when these can be practically manufactured.  In comparison,  we deal with a shell of fixed surface shape,  take bending forces into account
and use a more precise volume approximation, which leads to a non-convex problem. 

\begin{figure}[!htb]  
  \includegraphics[width=\columnwidth]{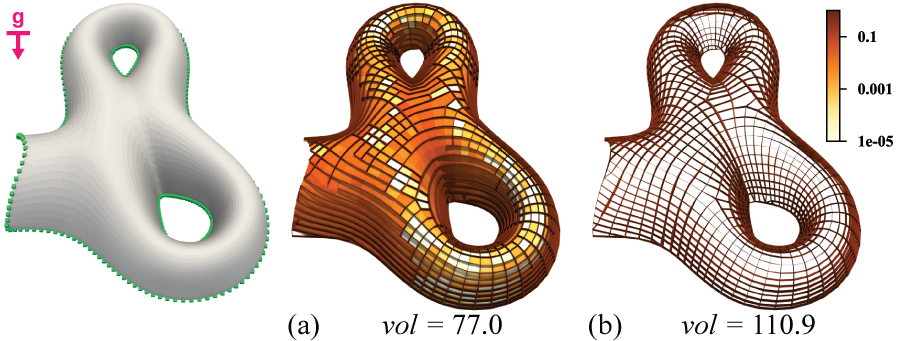}
  \centering
  \caption{Optimization results using different volume formula: (a) exact volume, (b) approximated convex volume.}
  \label{fig:compare-cvx-volume}
\end{figure}

 \cite{li2017rib} takes the approach closer to ours, in that it constructs a quadrangulation following a field to determine beam directions and takes bending into account in the elasticity model. 
 However, a narrow-beam approximation for the volume is used, and
 rather than using an optimized stress field as we do, the stress field of the non-optimized shell is used instead.  In some cases, the optimized field provides an advantage, although the results 
 are typically close as  Figure~\ref{fig:direction-opt} demonstrates. 
 More importantly, the volume approximation has a major effect as
 shown in Figure ~\ref{fig:compare-cvx-volume}, where we use one of the models we share with \cite{li2017rib}. In this case, the simpler volume approximation can increase the volume used by up to 44\%.
 On the other hand,  \cite{li2017rib} introduces I-beam optimization of beam cross-sections which we have not implemented, and may yield a substantial additional benefit. This technique can be easily combined with ours. Finally, \cite{li2017rib} does additional optimization of discrete beam directions, which can also be added to our method. 

\begin{figure}[!htb]  
  \includegraphics[width=\columnwidth]{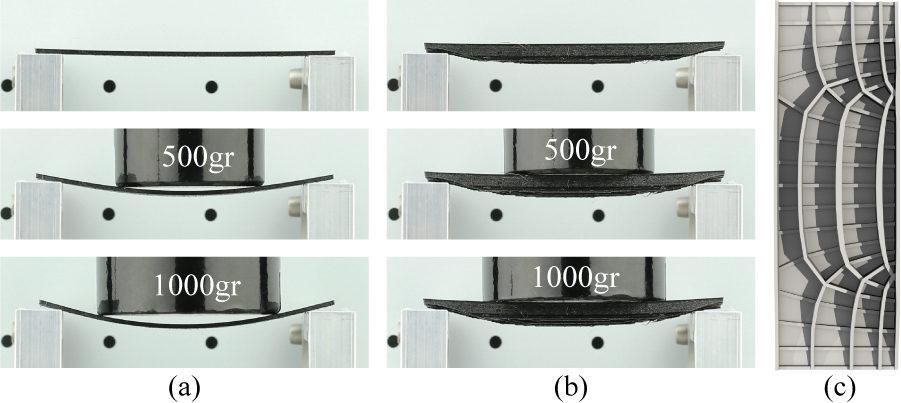}
  \centering
  \caption{(a) shelf based on original model (without optimization); (b) shelf printed from the optimized model; (c) optimized model.}
  \label{fig:experiment-bridge}
\end{figure}

\begin{figure}[!htb]  
  \includegraphics[width=\columnwidth]{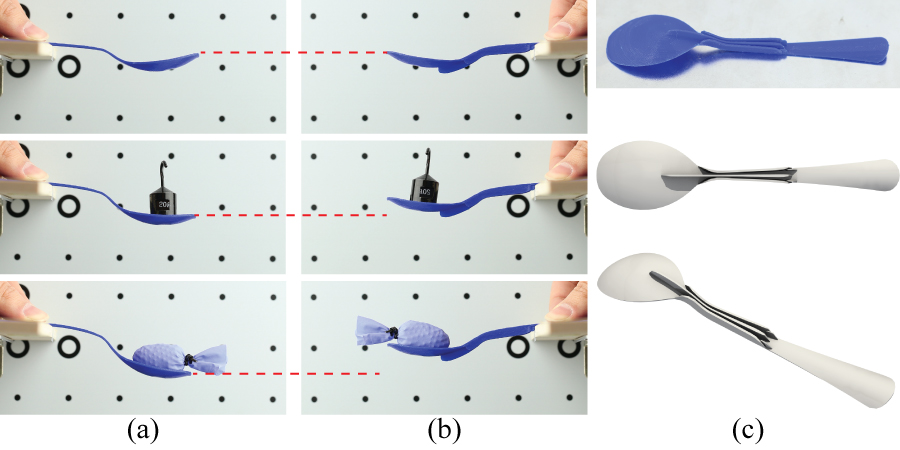}
  \centering
  \caption{(a) spoon based on original model (without optimization); (b) spoon printed from the optimized model; (c) optimized model.}
  \label{fig:experiment-spoon}
\end{figure}

\begin{figure}[!htb]  
  \includegraphics[width=\columnwidth]{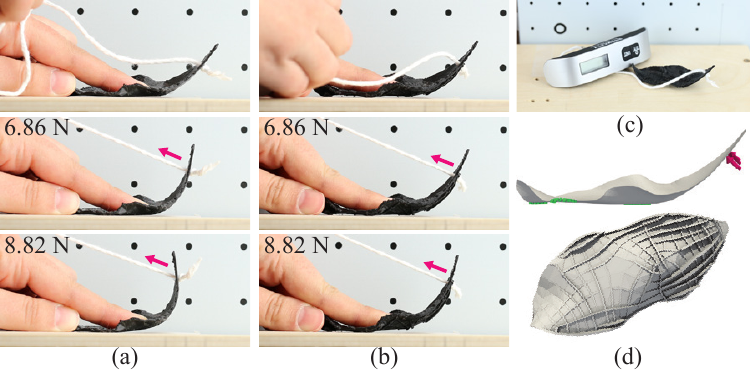}
  \centering
  \caption{(a) leaf based on original model (without optimization); (b) leaf printed from the optimized model; (c) experiment setup; (d) optimized model.}
  \label{fig:experiment-leaf}
\end{figure}

\paragraph{Experiments.} We have printed several simple structures to validate our optimization experimentally (Figures~\ref{fig:experiment-bridge},~\ref{fig:experiment-spoon},~\ref{fig:experiment-leaf}). Due to the highly approximate nature of the
physical model used, we did not attempt a quantitative match to the simulated values, but we did closely match the
values of the printed models. The comparison in all cases is between an optimized model and a uniform-thickness 
model of the same weight. Observed displacements in all cases differed by a factor more than 3, suggesting a similar 
difference in stress.

\begin{figure*}[htb]
\includegraphics[width=\textwidth]{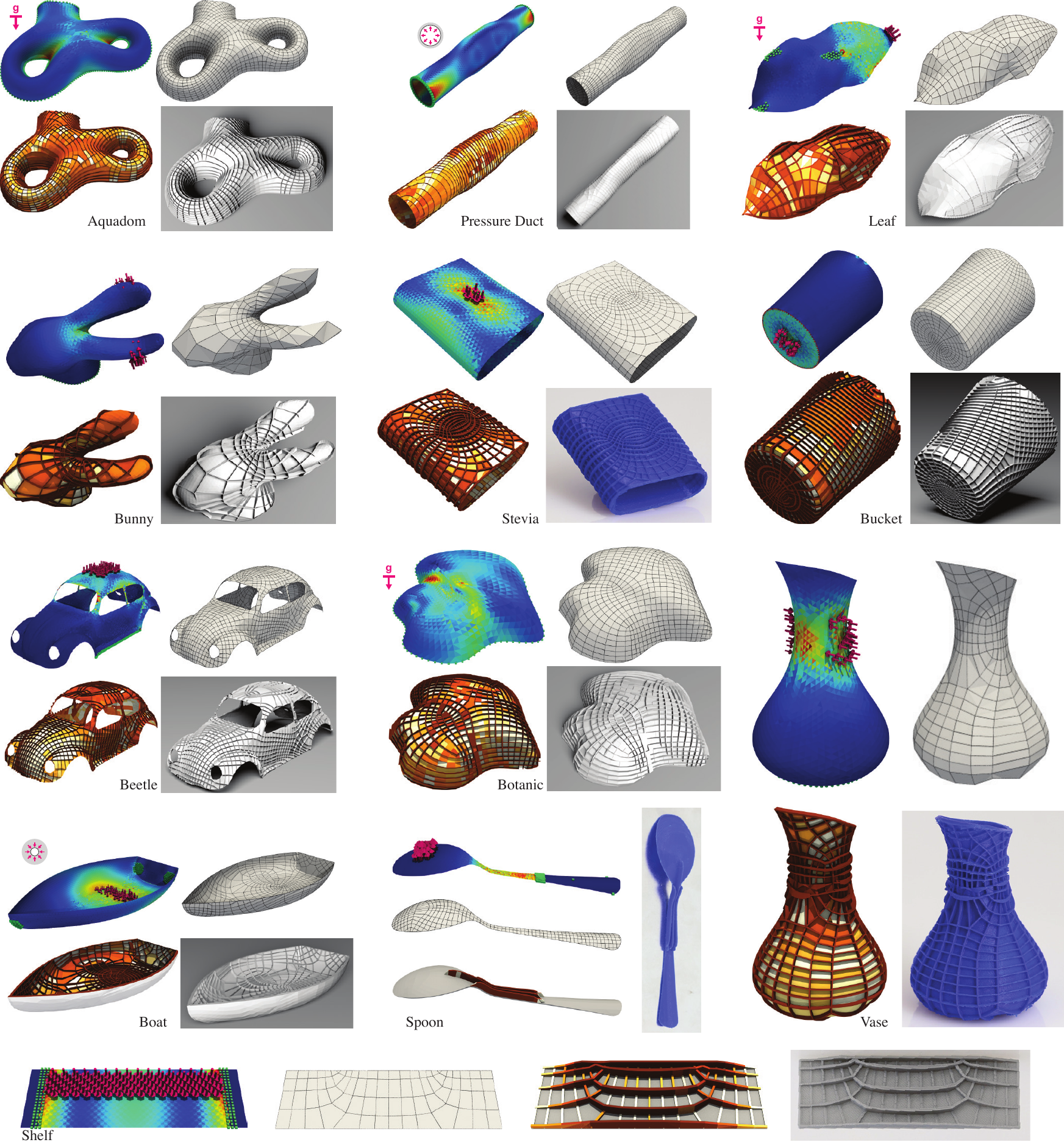}
\caption{Examples of structures obtained for a variety of shapes and loads. For each one, images show loads and initial stress distribution, quadrangulation, cell optimization (colored by thickness in logarithmic scale) and final geometry.
}
\label{fig:examples}
\end{figure*}

Finally, Figure~\ref{fig:examples} shows a set of optimized shells obtained for a variety of shapes using our method; In all cases we have preserved a minimal width/thickness beam to indicate the mesh edges, but the load
is carried by a relatively small number of beams; with shell shape constraints we found that unless the
thickness bound is set to very low values, even in tension areas ribs tend to appear. This is consistent with the
observation that with no thickness, the load carried by bending forces is maximized, if the goal is to reduce
the volume.  

\begin{figure}[!htb]  
  \includegraphics[width=\columnwidth]{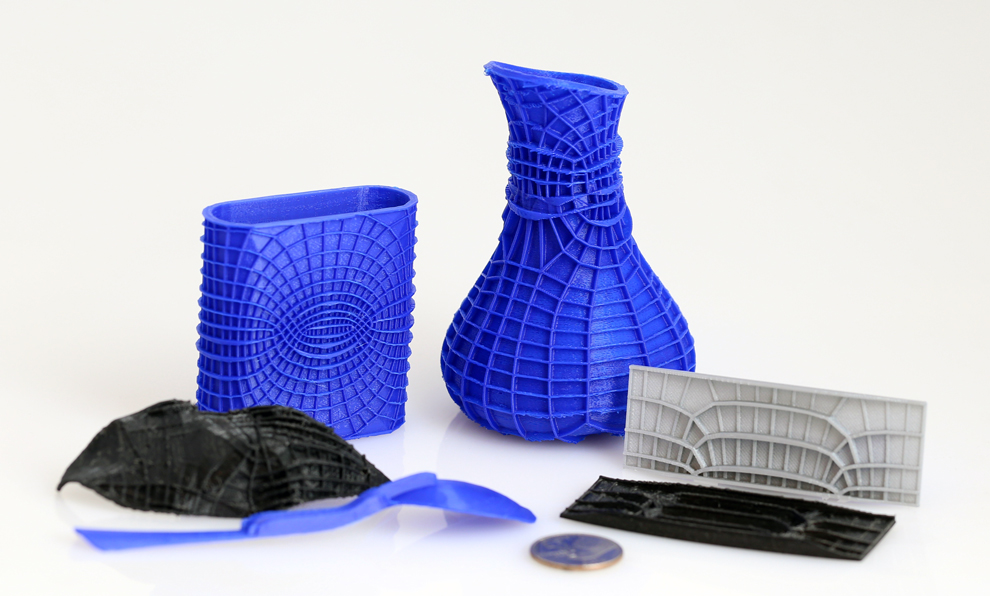}
  \centering
  \caption{Set of 3d printed models and a quarter (for scale).}
  \label{fig:3dprinted-objects}
\end{figure}

\begin{table}[htb!]
\vspace{-0.15cm}
\begin{tabular}{|l|llrrrrr|}
\hline
&d$(mm)$    &$|F|$ &$h$&$h^\text{s}$& $k$ &$s$&$t_\text{solve}(s)$\\
\hline
Aquadom  &85.60  &7742 &0.5         &0.005 &100 &0.2 &72.330\\
Beetle  &169.01  &7558 &1.2         &0.012 &50 &0.5 &35.999\\
Boat  &92.60  &9084 &1.0         &0.05 &100 &0.2 &33.067\\
Botanic  &42.59  &2152 &0.5         &0.01 &100 &0.5 &35.586\\
Bucket  &21.57  &7860 &0.3         &0.03 &50 &0.5 &45.744\\
Bunny  &106.00  &7471 &2.0         &0.9 &100 &0.25 &22.291\\
Duct  &460.58  &6932 &1.0         &0.01 &100 &0.2 &69.593\\
Leaf  &99.24  &2980 &2.0         &0.9 &46 &0.5 &1.954\\
Shelf  &88.54  &2400 &4.0         &0.5 &60 &0.5 &1.663\\
Spoon  &133.11  &6528 &4.0         &0.9 &29 &0.5 &4.129\\
Stevia  &90.68  &4840 &2.0         &0.9 &50 &0.5 &23.941\\
Vase  &69.64  &4798 &1.0         &0.45 &100 &0.5 &21.817\\
\hline
\end{tabular}
\caption{Statistics for the 3D models in Figure~\ref{fig:examples} and Figure~\ref{fig:teaser}: the bounding box diagonal $d$ (mm), the number of triangles $|F|$ in the input mesh, the maximum thickness $h$ (mm) of the beams, the constant shell thickness $h^\text{s}$ (mm), the number of iterations $k$ and step-size $s$ of the optimization solve, and the time $t_\text{solve}$ it took to complete all iterations.}
\label{tab:stats}
\vspace{-0.05cm}
\end{table}

\begin{table}[htb!]
\vspace{-0.15cm}
\begin{tabular}{|l|llrr|}
\hline
&$V^s (mm^3)$    &$V^b (mm^3)$ &$h_{eq}(mm)$& $c_{eq}/c$\\
\hline
Aquadom  &16.989  &90.335 &0.032               &1.601\\
Beetle  &170.128  &1287.227 &0.103             &5.739\\
Boat  &189.997  &418.539 &0.160                &7.099\\
Botanic  &9.134  &80.533 &0.098                &1.266\\
Bucket  &15.028  &8.325 &0.047                 &26.924\\
Bunny  &7270.428  &1134.806 &1.040             &1.229\\
Duct  &1075.645  &19777.817 &0.194             &2.105\\
Leaf  &2304.430  &659.348 &1.158               &1.105\\
Shelf  &1176.000  &1386.550 &1.090             &7.252\\
Spoon  &1544.960  &326.733 &1.090              &18.621\\
Stevia  &8329.664  &2568.409 &1.178            &1.198\\
Vase  &3505.556  &492.212 &1.026               &1.953\\
\hline
\end{tabular}
\caption{Performance for the 3D models in Figure~\ref{fig:examples}. We show the ratio $c_{eq}/c$ between the compliance $c$ of our results and the compliance $c_{eq}$ of a shell of constant thickness of the same weight. The fixed shell volume $V^s$, the support structure volume $V^b$, the thickness $h_{eq}$ of the equivalent-weight constant thickness shell. }
\label{tab:stats2}
\vspace{-0.05cm}
\end{table}


\section{Conclusions and Future Work}

In this paper, we have described a way to approximate and efficiently solve the problem of minimizing the weight of a
support structure for a shell.  The proposed method separates the construction
into three stages, with the first stage optimizing the field the beam directions must follow and creating a corresponding
quad-dominant mesh, the second stage creating a cell structure with optimized shape parameters, and the third stage creating an actual realization.

This makes it particularly flexible and allows one to
integrate a variety of additional user inputs and constraints, e.g., by modifying the field to change the truss directions, or
by adding beams in the second stage to support a connection to a separate object.
It is also very efficient, with optimization converging in a few iterations, and quite consistently.

While we use a highly simplified model for cell mechanics, the overall approach admits replacing this model with a more
advanced finite element formulation. In this case, it is likely that the local step would require numerical optimization;
however, as long as the model for a cell stays low-parametric, one is likely to be able to solve it efficiently. 

\paragraph{Limitations.} Our work has two main limitations: first, the formulation for the field optimization still has
restrictive low-volume and fixed-thickness assumptions. Based on our evaluation of field direction sensitivity of the final design,
we do not view this as a major limitation.  The second, more significant, limitation is the highly simplified model we used.
If exact results are needed, this model can be used to quickly obtain an initial result, which can then be refined using
a more advanced mechanical description of cells and shape optimization. 


\bibliographystyle{ACM-Reference-Format}
\bibliography{biblio} 

\appendix
\section{Convexity of  truss continuum optimization}
Expressing the functional in terms of entries of the stress matrix, using
$a = (\sigma_{11} + \sigma_{22})/2$, $b = (\sigma_{11}-\sigma_{22})/2$, 
and $c = \sigma_{12}$, we obtain


\begin{equation}
|\lambda_1(\sigma)| + |\lambda_2(\sigma)| = 2\max(a, \sqrt{b^2 + c^2})
\label{eq:eigenvals}
\end{equation}  
which verifies convexity of the energy.

\section{Dual of the continuum shell problem}
This derivation follows \cite{strang1983hencky}, which we include here for completeness. 

\begin{equation*}
  L(\sigma,\bu) = \frac{E}{\sigma_0} \int_\Omega |\lambda_1(\sigma)| + |\lambda_2(\sigma)|dA + \int_\Omega \bu^T \mathrm{div} \sigma dA - \int_\Omega \bbf^T \mu dA.
\end{equation*}
Integrating by parts, and assuming either free boundary $\partial_n \bu = 0$
or fixed boundary $\bu = 0$, we obtain

\begin{equation*}
L(\sigma,\bu) = \frac{E}{\sigma_0} \int_\Omega |\lambda_1(\sigma)| + |\lambda_2(\sigma)|dA - \int_\Omega \eps(\bu): \sigma dA - \int_\Omega \bbf^T \bu dA.
\end{equation*}

To obtain the dual we need to minimize over all possible $\sigma$ in a coordinate system aligned with $\eps$ the expression

$$\frac{E}{\sigma_0}( |\lambda_1(\sigma)| + |\lambda_2(\sigma)|) -  \lambda_1(\eps)\sigma_{11} -
\lambda_2(\eps)\sigma_{22}.$$
The minimum of this expression is $-\infty$, if $|\lambda_i(\eps)| > 1$, for either $i=1$ or $i=2$;
otherwise, it is zero; this leads to the dual problem \eqref{eq:dual-continuum}.

\section{Proof of the properties of $V(z)$.}
\label{sec:proof}
We drop the cell superscript $c$ in this proof.
We assume that $z_1 \leq z_2 \leq z_3$. 
By the constraints of the problem, $y_1+y_2 +y_3 \leq 1$,  $0 < z_{min} \leq z_i$.

$V(y,h)$ \eqref{eq:tri-volume} is a quadratic function of $y_i$ and 
linear in $h_i$; once we express it in terms of $z_i$, to eliminate 
the constraint between $h_i$ and $z_i$, it becomes a cubic function  of $z_i$, with 
Hessian 

$$
\scriptsize
H(z)= \left[ 
\begin {array}{ccc} 
A & -k_1 g^b_2&    -k_1 g^b_3\\
\noalign{\medskip}
-k_1 g^b_2 & B &-k_2 g^b_3\\
\noalign{\medskip}-k_1 g^b_3 & -k_2 g^b_3 &  -1/2g^b_3  \left( g^t_3  +3 k_3 \right)
\end {array} \right]
$$
where $k_i = g^t_i+ 2g^b_i z_i = dw_i/dz_i$, 
$B = -1/2g^b_2\, \left( g^t_2+3 k_2+2 g^t_3+2 k_3 \right)$ and
$A = -1/2 g^b_1\, \left( g^t_1  + 3 k_1+2 g^t_2 + 2 k_2+ 2 g^t_3 + 2 k_3 \right)$.  

A direct evaluation shows that  
$v^T H(z) v$ evaluated for 
$v = [1,0,0]$ is negative for positive 
$z_i$ and $g^b_1\neq 0$. Similar is true
for $v = [0,1,0]$ and $v = [0,0,1]$. 
We conclude if $g_i^b \neq 0$ for some $i$, 
any critical point in the interior of the domain is a maximum or a saddle, so there are no minima in the interior. 
If all $g_i^b$ are zero, the volume is 
a linear function of $z_i$, and the 
optimum is also on the boundary. 

The constraint $y_1 + y_2 + y_3 \leq 1$ 
defines, if $g^i_b \neq 0$ for some $i$ 
a quadratic surface in $z_i$ coordinates, and 
the constraints $z_1 \leq z_2$, $z_2 \leq z_3$,
$z_1 \geq z_{min}$ define 3 halfspaces. 
Their intersection is a tetrahedron with 
a curved face on the ellipsoid. 
The faces of the feasible domain correspond to one of the inequality constraints becoming an equality.  Similarly, 
by a direct calculation, substituting either 
$z_1 = z_{min}$, $z_2 = z_1$ or $z_2 = z_3$ into $V(z)$ and computing $2 \time 2$ Hessians, we observe that these are not positive definite, hence  there can be no solution on faces corresponding to the linear constraint. Finally, the same fact holds for pairs of constraints that define edges of the constraint domain, e.g., $z_1 = z_{min}$  and $z_2 = z_1$. We conclude that either the solution is on the face corresponding to the constraint 
$y_1 + y_2 + y_3 = 1$ or at the only vertex not on this face of the feasible domain, $z_i = z_{min}$,  $i=1,2,3$ (case 1 of the proposition).  If the minimum satisfies $y_1 + y_2 + y_3 = 1$,  it is either in the interior (case 2), or on one of the three edges of the boundary of that face (cases 3-5). 

\section{Discrete bending strain for beams}

The common normal $\bn_i$ at $i$ is defined as $$\bn_i = \sum_{\ell \in N(i)} \be_{i\ell} \times \be_{i\,next(\ell)}$$
where $next(\ell)$ is the edge following $\ell$ in CCW order around $i$.

At each vertex, we define a normal $\bn_i$, as the average of cross-products of pairs of incident edges.
For an edge $\be_{il}$,  incident at a vertex $i$, let $\be_{im}$ be the next CCW edge,  and $\be_{ip}$ be the previous edge. We define 
$$\bq_{lm} = \be_{il} \times \be_{im}$$ the (unscaled) normal to the triangle formed by $i,l,m$.
Then $\bn_i = \sum_{l,m \in N(i)} \bq_{lm}$, and  $\bhn_i = \bn_i/|\bn_i|$.

We express the bending strain in the form $D^b_{ij} u$, where  $u$ is the vector of all vertex displacements.  Define $\Delta \be_{ij} = \bu_j - \bu_i$.
Then $\Delta \bq_{lm} = \Delta \be_{il} \times  \be_{im}  + \be_{il} \times \Delta \be_{im}$. The part of $\Delta \bq_{lm}$ affecting $\Delta \bhn_i$ is the part perpendicular to $\bhn_i$: 

$\Delta q_{lm}^\perp = (I - \bhn_i \bhn_i^T) \Delta \bq_{lm} = P_i \Delta \bq_{lm}$, 
where $P_i$ is the matrix $(I - \bhn_i \bhn_i^T)$. By substitution into the expression for $n$ we get:
$$\Delta \bhn_i = \frac{1}{|\bn_i|} P_i \sum_{l \in N(i)}  g_l \times (\bu_l - \bu_i)$$ 

where $g_l$ is defined as follows: 
\begin{itemize}
\item $\bg_l = \be_{ip} -\be_{im}$, if both $p$,$m$ are in $N(i)$, 
i.e. $\be_{il}$ is not on the boundary. 
\item $\bg_l = \be_{ip}$, if $m \not\in N(i)$.
\item $\bg_l = -\be_{im}$, if $p \not\in N(i)$.
\end{itemize}

For a vector $\ba = [a_x, a_y, a_z]$, let $R(a)$ be the infinitesimal rotation matrix about $a$:
$\left(\begin{array}{rrr} 0 &-a_z &a_y\\ a_z& 0 & -a_x\\-a_y& a_x & 0\end{array}\right)$.

Then 
$$\Delta \bhn_i =  \sum_{l \in N(i)}  M^i_{l} (\bu_l - \bu_i)$$ where $M^i_{l} = \frac{1}{|\bn_i|} P_i R(\bg_l)$ is a $3\times 3$ matrix. 

 $$\eps^b[ij]  = h \left( \sum_{l \in N(j)} (\bd_{ji}^l)^T (\bu_l-\bu_j) + \sum_{l \in N(i)} (\bd_{ij}^l)^T (\bu_l-\bu_i) \right)$$
where $\bd_{ij}^l = -(M^i_{l})^T \bhe_{ij}/l_{ij}$, a vector of length 3. From this expression, we can immediately obtain $D_{ij}$.


\end{document}